\documentclass{aastex63}

\received{March 15 2020}
\revised{May 13 2020}
\accepted{May 26 2020}

\submitjournal{PSJ}

\shorttitle{Dynamical History of the Uranian System}
\shortauthors{{\'C}uk et al.}

\begin{document}

\title{Dynamical History of the Uranian System}

\correspondingauthor{Matija {\'C}uk}
\email{mcuk@seti.org}

\author{Matija {\'C}uk}
\affil{SETI Institute \\
189 North Bernardo Ave, Suite 200 \\
Mountain View, CA 94043, USA}

\author{Maryame El Moutamid}
\affiliation{Carl Sagan Institute and Department of Astronomy \\
Cornell University  \\
326 Space Sciences Building \\
Ithaca, NY 14853, USA}

\author{Matthew S. Tiscareno}
\affil{SETI Institute \\
189 North Bernardo Ave, Suite 200 \\
Mountain View, CA 94043, USA}

\begin{abstract}
We numerically simulate the past tidal evolution of the five large moons of Uranus (Miranda, Ariel, Umbriel, Titania, and Oberon). We find that the most recent major mean-motion resonance (MMR) between any two moons, the Ariel-Umbriel 5:3 MMR, had a large effect on the whole system. Our results suggest that this resonance is responsible for the current 4.3$^{\circ}$ inclination of Miranda (instead of previously proposed 3:1 Miranda-Umbriel MMR), and that all five moons had their inclinations excited during this resonance. Miranda experienced significant tidal heating during the Ariel-Umbriel 5:3 MMR due to its eccentricity being excited by Ariel's secular perturbations. This tidal heating draws energy from shrinking of Miranda's orbit, rather than Ariel's outward evolution, and can generate heat flows in excess of 100~mW m$^{-2}$, sufficient to produce young coronae on Miranda. We find that this MMR was followed by a sequence of secular resonances, which reshuffled the moons' eccentricities and inclinations. We also find that the precession of Oberon's spin axis is close to a resonance with the precession of Umbriel's orbital plane, and that this spin-orbit resonance was likely excited during the Ariel-Umbriel 5:3 MMR. After the exit from the MMR, subsequent Ariel-Umbriel secular resonance and Oberon-Umbriel spin-orbit resonance may be able to explain the current low inclinations of Ariel and Umbriel. The age of Miranda's surface features tentatively suggests Uranian tidal $Q=15,000-20,000$, which can be further refined in future work.   
\end{abstract}


\section{Introduction} \label{sec:intro}

Uranus has five large satellites (Miranda, Ariel, Umbriel, Titania, and Oberon) as well as a number of smaller ``ring moons`` closer to the planet and irregular satellites at large distances. The orbital planes of the five major moons are relatively close to the equator of Uranus (Table \ref{table1}), indicating that these moons originated in a disk of material around Uranus after the planet itself formed.

\begin{table}[!ht]
\begin{center}
\caption{Sizes and mean orbital parameters of Uranus's five large regular satellites. Orbits of Ariel-Oberon \citep[credited to ][]{las87}, as well as Miranda-Oberon masses \citep[credited to ][]{jac07} and radii \citep[credited to ][]{tho88} were retrieved from the Jet Propulsion Laboratory's Solar System Dynamics website ssd.jpl.nasa.gov on 12/10/16. The orbit of Miranda is taken from \citet{sho06}. Semimajor axes are scaled using the reference value of Uranus's equatorial radius, 25,559~km, and inclinations are measured relative to Uranus's equator.\label{table1}}
\begin{tabular}{|l|c|c|c|c|c|}
\hline\hline
Moon & Mass & Radius & Semimajor & Eccentricity & Inclination \\ 
\ & ($10^{20}$~kg)& (km) & axis ($R_U$) & & ($^{\circ}$)\\ 
\hline\hline
Miranda & 0.66 & 236 & 5.080 & 0.0006 & 4.449 \\
\hline
Ariel & 13.53 & 579 & 7.469 & 0.0012 & 0.041 \\
\hline
Umbriel & 11.72 & 585 & 10.407 & 0.0039 & 0.128 \\
\hline
Titania & 35.27 & 789 & 17.070 & 0.0011 & 0.079\\
\hline
Oberon & 30.14 & 761 & 22.830 & 0.0014 & 0.068\\
\hline
\end{tabular}
\end{center}
\end{table}

The largest departure from the circular and equatorial orbits among the five major moons is the four-degree inclination of Miranda, which implies strong dynamical perturbation of Miranda's orbit at some point after its formation. Other indication of past orbital excitation is a dramatic pattern of surface features on Miranda \citep{smi86}, as well as indications of past resurfacing on Ariel \citep{ple87}, despite these moons' present-day small orbital eccentricities and correspondingly low tidal heating rates. The most common explanation for past perturbations is that a pair of moons encountered mutual mean-motion resonances due to differential tidal evolution. 

Currently, none of the five major moons are in any mean-motion resonance with each other, although Miranda's mean motion is affected by its relative proximity to a three-body resonance with an argument $\lambda_M-3 \lambda_A +2 \lambda_U$ \citep[where $\lambda$ is mean longitude, and subscripts refer to the moons' initials;][]{gre75,jac14}. However, since Ariel's outward tidal evolution is expected to be faster than that of Miranda \citep[Eq.4.167 from ][, assuming no frequency dependence of Uranus's tidal parameters]{md99}, the moons will only cross the exact three-body resonance in the future, so this resonance should have had little effect on the satellite orbits in the past.

\begin{figure*}
\epsscale{.6}
\plotone{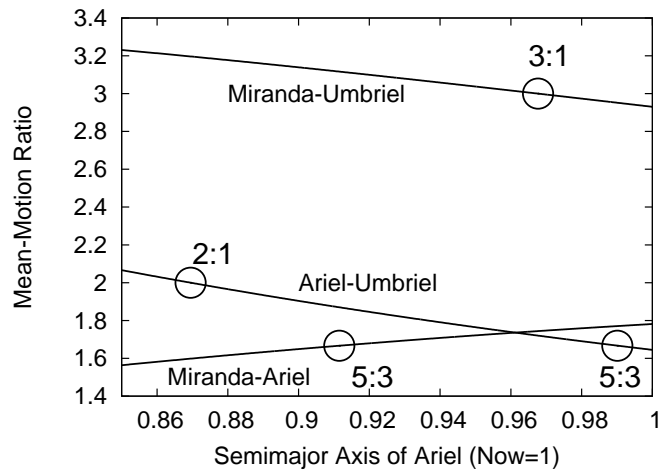}
\caption{Past mean-motion ratios between Miranda, Ariel and Umbriel plotted against the semimajor axis of Ariel (as the moon with the fastest tidal evolution). Major resonances are labeled using open circles. This evolution of period ratios were calculated using the Eq. 4.167 from \citet{md99}, with satellite parameters from Table \ref{table1}. We assumed frequency-independent tidal $Q$ for Uranus. Plot design inspired by \citet{pea88}.\label{peale}}
\end{figure*}

The most recent major resonance that the moons of Uranus should have encountered in the course of their tidal evolution is the 5:3 mean motion resonance (MMR) between Ariel and Umbriel (Fig. \ref{peale}). This is a second order resonance, so we would expect both eccentricities and inclinations of the two moons to be affected by resonance passage \citep[][, section 8.8.2]{md99}. However, the only study of this resonance to date employed a semi-analytical approximation that assumed planar orbits, ignoring any effects on inclinations \citep{tw1}. Three decades later, direct numerical integration is feasible and the Arel-Umbriel 5:3 MMR passage clearly needs to be revisited.

Since the 5:3 Ariel-Umbriel MMR was not thought to have affected Miranda, the established hypothesis for the origin of Miranda's inclination is the past 3:1 Miranda-Umbriel MMR, which should have predated the Ariel-Umbriel 5:3 MMR \citep{tw2, mal90}. If Miranda was originally on a low-inclination orbit, capture into the pure inclination subresonance \citep[$i_M^2$ in the nomenclature of ][]{md99} is very likely, and the inclination of Miranda would subsequently increase with time. It is thought that the Miranda-Umbriel 3:1 MMR was broken at Miranda's inclination of $4.3^{\circ}$ by a secondary resonance \citep{tw2, mal90, ver13}. 

An even earlier Ariel-Umbriel 2:1 MMR crossing probably never happened, as this resonance would be very difficult to break once established \citep{tw3}. This enabled \citet{tw3} to put the limit of $11,000 < Q < 39,000$ on the tidal quality factor of Uranus, by assuming that the system is 4.5~gyr old and that the 3:1 Miranda-Umbriel MMR was crossed but the Ariel-Umbriel 2:1 MMR was not (Fig. \ref{peale}). 

In this paper, we aim to use direct numerical simulations to revisit the Ariel-Umbriel 5:3 MMR, which was previously only studied using analytical approximations \citep{tw1}. This should have been the most recent resonance in the Uranian system, and therefore the effects of this resonance should be most clearly visible in the current properties of the moons. Additionally, \citet{pet15} find that the ``classic'' picture of the past of the Uranian satellites fails to explain apparent past tidal heating of Ariel, as has already been long known for Miranda \citep{pea88}. It is possible that direct numerical simulations can identify previously overlooked sources of orbital excitation and tidal heating, and place new constraints on the tidal properties of Uranus and its moons.  

In the following sections, we describe the different kinds of resonant behavior that we find to have played a role in the evolution of the Uranian satellites.  According to our narrative, substantial shaping of the system by the Ariel-Umbriel 5:3 MMR (Section \ref{sec:mmr}) was followed by re-distribution of eccentricity and inclination among the moons by secular resonances (Section \ref{sec:secular}), and inclinations may have been damped by a spin-orbit resonance between Oberon’s obliquity precession and a secular mode (Section \ref{sec:spin}).  The modern eccentricity of Ariel may have arisen even later, due to a three-body resonance involving Ariel, Umbriel, and Titania (Section \ref{sec:3br}).  General discussion is given in Section \ref{sec:dis}, and conclusions are presented in Section \ref{sec:con}.

\section{Ariel-Umbriel 5:3 MMR} \label{sec:mmr}

In this work we primarily use numerical integrator {\sc simpl} \citep{cuk16}, which uses a mixed-variable symplectic \citep{wis91} algorithm developed by \citet{cha02}. This program integrates full equations of motion, and models perturbations from the planet's oblateness, other moons, the Sun and the planets, and also approximates both planetary and satellite tides. For a full description of {\sc simpl} we refer the reader to \citet{cuk16}. In this paper we assume everywhere that $Q/k_2$ for both Uranus and the satellites is not a function of frequency. In all of our integrations we included the solar perturbations, but in this work no planetary perturbations were included. For each integration we will specify which $Q/k_2$ was used for Uranus; enhancing the tidal response of Uranus is the only method we use for accelerating our runs. 

\begin{figure*}
\epsscale{.6}
\plotone{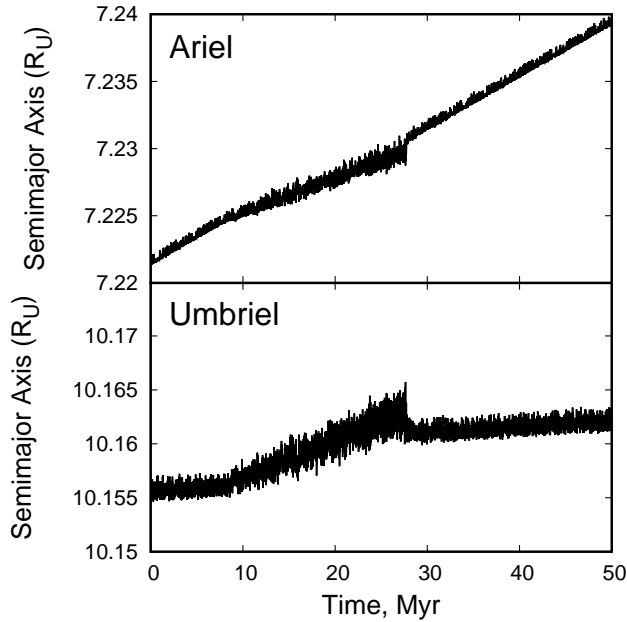}
\caption{Evolution of semimajor axes of Ariel and Umbriel in a simulation of their 5:3 MMR crossing. In this simulation we used $Q/k_2= 4 \times 10^4$ for Uranus (i.e. evolution was accelerated about 10 times over our nominal case). The resonance is established about 10 Myr into the simulation and lasts for about 20 Myr, at which point it spontaneously breaks.\label{accel_a}}
\end{figure*}

In most of our simulations, we obtain initial conditions by taking present-day satellite orbits, and modify them in two ways: we set the inclination of Miranda to be about $0.1^{\circ}$, and we change Ariel's semimajor axis so that it is immediately interior to the 5:3 MMR with Umbriel (in Fig. \ref{accel_a}, Ariel is initially only 0.04\% interior to the exact commensurability). Different simulations of this resonant crossing featured in this paper are obtained by varying tidal parameters of Uranus and the moons, as well as the exact placement of Ariel interior to the resonance. 

The tidal properties of Uranus can be constrained only once we have a unique solution for the satellites' dynamical history, but it is clear from prior work by \citet{tw3} that the tidal quality factor $Q$ of Uranus must be in the approximate $10^4 < Q < 10^5$ range. We also use tidal Love number $k_2=0.1$ for Uranus, which is a conveniently round quantity and very close to the results of \citet{gav77}. We also assume that the tidal $Q$ of Uranus is not frequency-dependent. In practice, since the timing of the Ariel-Umbriel 5:3 MMR crossing we study here is determined by Ariel's tidal evolution, by ``tidal $Q$'' we mean ``tidal $Q$ at Ariel's synodic frequency.'' Whenever we discuss accelerated tidal evolution, the nominal ``realistic" tidal dissipation within Uranus is taken to correspond to $Q=4 \times 10^4$ (with $k_2=0.1$), which is roughly the smallest tidal $Q$ for which Miranda did not cross the 3:1 MMR with Umbriel according to \citet{tw2}. This ``nominal'' value of Uranus tidal $Q$ has been selected at the very beginning of this project (when our preliminary runs indicated excitation of Miranda's inclination in the Ariel-Umbriel 5:3 resonance), and was used as a guide to what tidal evolution rates should be considered realistic. This nominal assumption should not be confused with the tidal $Q$ of Uranus derived from our results, which is presented and discussed in Section \ref{sec:dis}. 

Tidal response of the satellites is even harder to estimate. Since our integrator uses only the ratio $Q/k_2$ to calculate tidal accelerations rather than treating $Q$ and $k_2$ separately, we will set the satellites' $Q=100$ for all simulations \citep[a common choice for solid bodies;][, section 4.9]{md99}, and only vary their tidal Love numbers. According to \citet{md99}, when treated as solid icy bodies, Miranda and Ariel should have approximate tidal Love numbers of $k_{2, M}=0.001$ and $k_{2,A}=0.01$. More rigid response (possibly due to very low temperatures) could lead to a lower $k_2$, while a subsurface ocean could result in a very large Love number \citep[Enceladus apparently has $Q/k_2 \simeq 100$;][]{lai12}. Therefore, in different simulations we will use tidal Love number $k_2$ for Ariel ranging from $k_2=0.001$ (very cold interior) to $k_2=1$ (global ocean?). 

Integrations of the five major Uranian moons using {\sc simpl} on standard workstation processors require about a day of computation time for 2 Myr of simulation time. This makes simulations longer than 50-100 Myr rather cumbersome as they require multiple months of processor time. Therefore we ran a mixture of accelerated and realistic simulations, with the accelerated simulations enabling us to explore the phase space, while the few realistic simulations act as a check that the acceleration of tidal forces is not introducing major artifacts.  

\begin{figure*}
\plotone{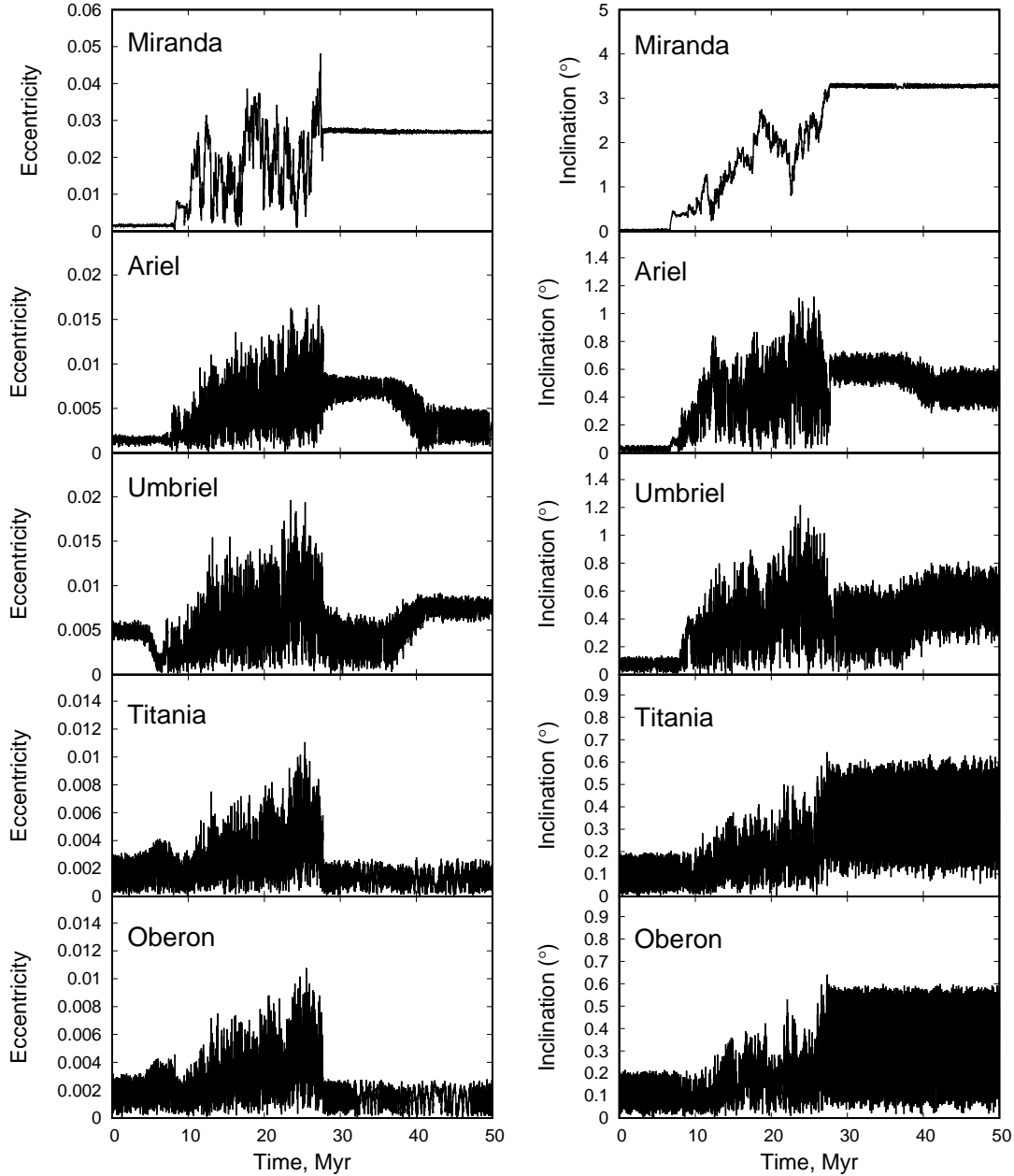}
\caption{The evolution of eccentricities and inclinations of Uranus's major moons in the simulation shown in Fig. \ref{accel_a}. All  five moons' eccentricities and inclinations are excited during the Ariel-Umbriel 5:3 MMR (9--28~Myr in the simulation). There is also a secular resonance between modes 2 and 3 (in practice, Ariel and Umbriel) at 36--42 Myr, and a short-lived secular resonance between mode E3 (associated with eccentricity of Umbriel) and E4 (contributing to eccentricities of Titania and Oberon) at 5--6 Myr.\label{accel_ei}}
\end{figure*}

Fig. \ref{accel_a} shows semimajor axes of Ariel and Umbriel in a simulation that has been accelerated about 10 times over our nominal tidal evolution (i.e. we used $Q/k_2=4 \times 10^4$ for Uranus). Here the moons were assumed to be very rigid and non-dissipative ($Q/k_2=10^6$ for Miranda, $Q/k_2=10^5$ for other moons). Fig. \ref{accel_a} shows Ariel and Umbriel spending just under 20 Myr in their 5:3 resonance, before the resonant lock is broken. Figure \ref{accel_ei} shows the eccentricities ($e$) and inclinations ($i$) of all five moons in the same simulation. Contrary to the usual expectation for a two-body MMR, the $e$ and $i$ of non-participating moons are also excited by the resonance. This is due to strong secular coupling  between the satellites \citep{las87}, which redistributes $e$ and $i$ between the moons' orbits. The amount of orbital excitation is inversely proportional to the mass of the satellite: Miranda is most strongly affected, followed by Ariel and Umbriel, with Titania and Oberon having lowest eccentricities and inclinations.

\begin{figure*}
\plotone{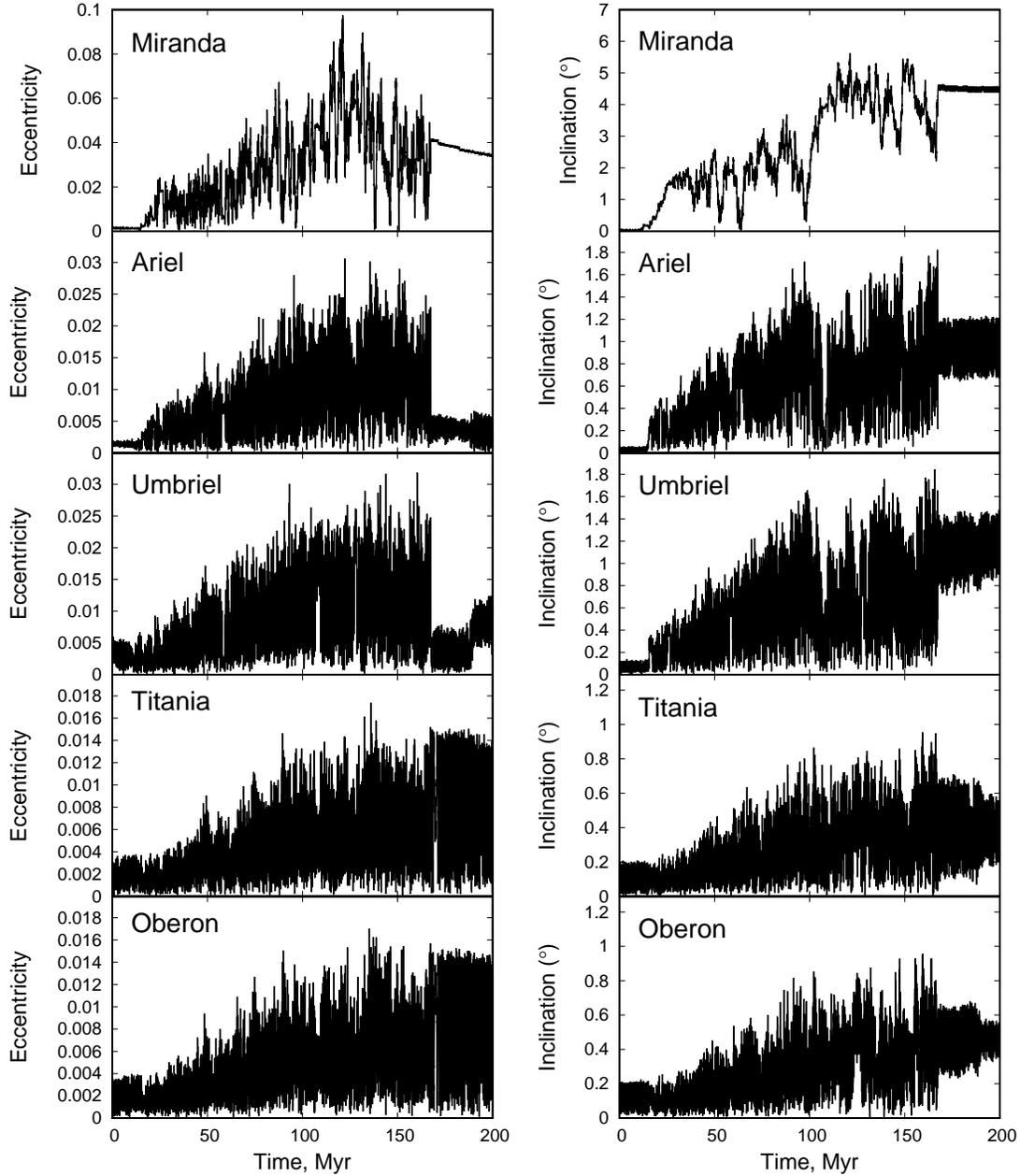}
\caption{Eccentricities and inclinations of the five major moons during a simulation of Ariel-Umbriel 5:3 MMR using $Q/k2=1.33 \times 10^5$ for Uranus. This rate of tidal evolution is three times faster than our nominal value, and is at the higher limit of the range allowed by dynamical constraints (Section \ref{sec:dis}). The Ariel-Umbriel resonance happens during the 15--165 Myr interval, and is followed by secular resonance between modes 3 (associated with Umbriel and, to a lesser extent, Ariel) and 4 (associated with Titania and Oberon) at 185 Myr.\label{long}}
\end{figure*}

In order to make sure that the dynamical picture we see is not an artifact of the enhanced tidal accelerations in our model, we also conduced a long-duration simulation of the Ariel-Umbriel 5:3 resonance crossing using more realistic tidal accelerations. Figure \ref{long} shows an evolution that was accelerated approximately 3 times over nominal case ($Q/k_2= 1.33 \times 10^5$ for Uranus). The satellites were still considered rigid, although somewhat more dissipative ($Q/k_2=10^5$ for Miranda, $Q/k_2=10^4$ for other moons). The $e$ and $i$ attained by all five moons in Fig. \ref{long} are comparable to those shown in Fig. \ref{accel_ei}, and possibly even slightly higher. After running a number of simulations using a range of tidal parameters for the planet and the satellites, we conclude that the excitation of the all five satellite orbits, followed by resonance breaking, is the guaranteed outcome when the tidal damping within the moons, especially Ariel, is low ($Q/k_2 \gtrsim 10^4$). 

\begin{figure*}
\plotone{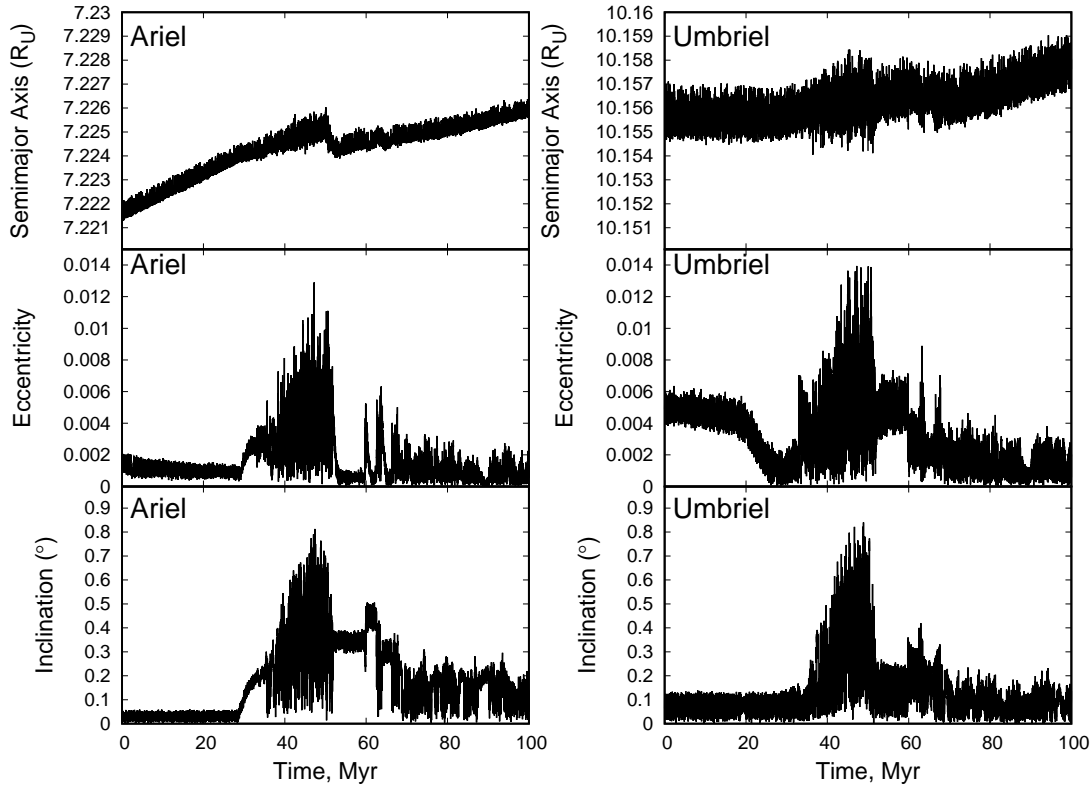}
\caption{Simulation of the Ariel-Umbriel 5:3 MMR during which we greatly increased the rate of tidal dissipation within Ariel after 50~Myr, as a proxy for internal melting. At the start of the simulation, we used $Q/k_2=2 \times 10^5$ for Uranus and $Q/k_2=10^4$ for Ariel. The resonance is established at about 30 Myr and at 50 Myr we changed Ariel's tidal response to $Q/k_2=100$ \citep[similar to that of Enceladus;][]{lai12}. At first Ariel drops out of resonance due to inward tidal evolution caused by intense eccentricity tides, but the resonance reestablishes itself at about 65~Myr and continues to the end of the simulation. Eccentricities and inclinations of Ariel and Umbriel are now kept low ($e\leq0.002$, $i\leq 0.2^{\circ}$) by Ariel's tidal dissipation.\label{damp}}
\end{figure*}

\begin{figure*}
\plotone{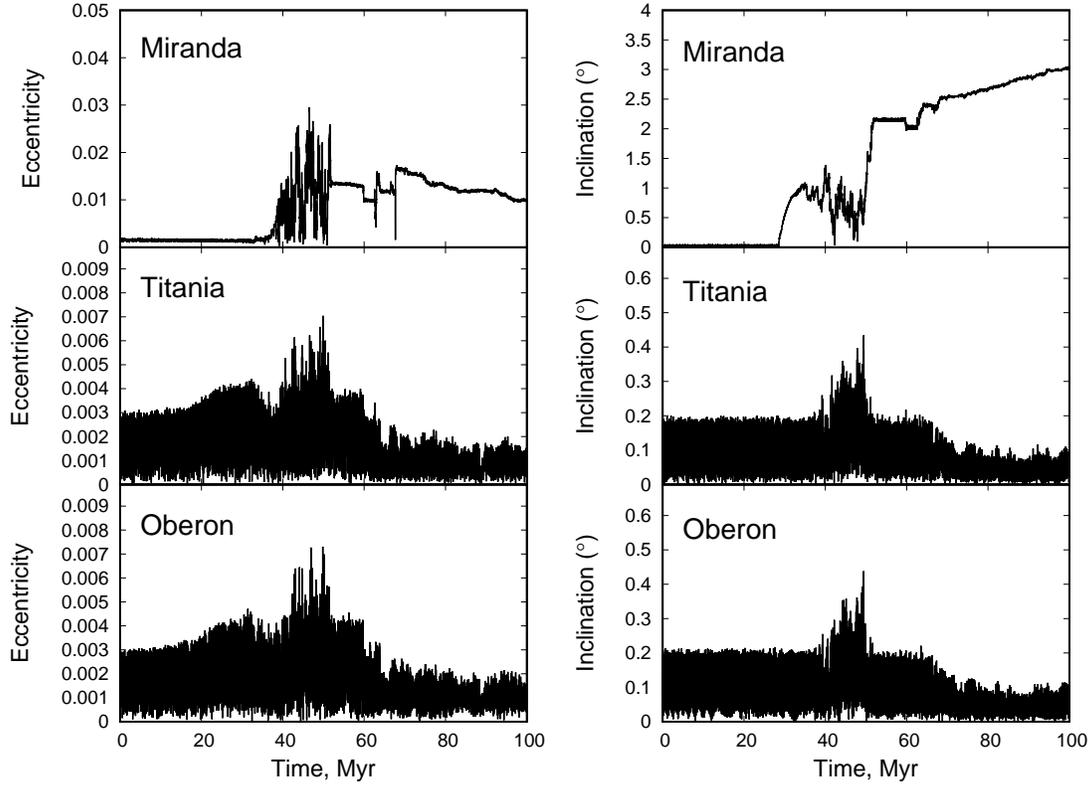}
\caption{Eccentricities and inclinations of Miranda, Titania, and Oberon in the simulation  shown in Fig. \ref{damp}. The eccentricities and inclinations of Titania and Oberon are both suppressed after the onset of Ariel' strong tidal dissipation. The eccentricity of Miranda becomes much less chaotic and its long-term evolution is dominated by tidal damping (within Miranda). The inclination of Miranda continues to grow apparently indefinitely.\label{damp2}}
\end{figure*}

To illustrate the effect of strong tidal damping within Ariel, in Fig. \ref{damp} we show a simulation in which we sharply reduced the $Q/k_2$ of Ariel after about 20 Myr in the resonance (i.e. 50 Myr into the this simulation). In this simulation we used $Q/k_2=2\times 10^5$ for Uranus (i.e. accelerated by about a factor of two). Up to 50 Myr into the simulation, the moons' tidal parameters were $Q/k_2=10^5$ (Miranda) and $Q/k_2=10^4$ (other moons). At 50 Myr, we instantaneously switched tidal dissipation within Ariel to $Q/k_2=100$, with the intention of approximating large-scale melting. The eccentricity and inclination of Ariel and Umbriel drop instantaneously, and Ariel temporarily leaves the resonance. At 60--70 Myr the resonance is re-established, but now with much less excited orbits for all five moons (Figs \ref{damp} and \ref{damp2}, middle and bottom panels). The exception is the inclination of Miranda, which keeps growing over time, while Miranda's eccentricity appears to be slowly damped by its own satellite tides (Fig. \ref{damp2}, top panels). 

The possibility that the moons of Uranus once inhabited a damped resonant state is somewhat attractive, as it would explain the low inclinations of the four larger moons, while still allowing Miranda to attain a large inclination, as observed. However, in numerous simulations that we ran we never saw any sign of Ariel and Umbriel being able to leave the resonance while at low eccentricities. In that respect we are in agreement with \citet{tw1}, who find that eccentricities of about $e=0.02$ are needed to exit the resonance. While it is impossible to prove the negative using numerical simulations, we have accumulated several Gyr of simulations of damped Ariel-Umbriel 5:3 resonance, using a variety of system parameters, and no resonance breaking was observed. While it is in principle possible that there is some aspect of the problem we have missed, we must conclude at this point that Ariel and Umbriel cannot leave their 5:3 MMR if their eccentricities are strongly damped. 

We also considered the possibility that the resonance may have been broken by an impact. The most advantageous impact would be a head-on impact onto Umbriel, which would move it to a lower orbit and out of the resonance. Alternatively, a resonance-breaking impact onto Ariel would have to be in the direction of orbital motion, reducing the relative velocity between the impactor and the moon (impacts that make the two moons' orbits converge would lead them to soon re-enter the resonance). By varying Umbriel's initial $a$ in many short simulations, we find that the relative change of Umbriel's semimajor axis required to exit the resonance is $\Delta a/a=2 \times 10^{-4}$. If we assume orbital velocity of 5~km/s for Umbriel and an impact velocity of 10 km/s (exactly opposite the orbital motion), we find that the impactor should be $0.5 \times 10^{-4}$ the mass of Umbriel, or about 25 times smaller in diameter. A~50 km impactor would make a 500+~km crater on Umbriel\footnote{Using Keith Holsapple Web-based crater size calculator: http://keith.aa.washington.edu/craterdata/scaling/index.htm}, which should be geologically young. While there are indications of a large basin on Umbriel \citep{tho88}, it appears to be geologically old, rather than recent. More problematic is the probability of such an impact: \citet{zah03} estimate that the rate of formation of 500~km craters on Umbriel is $2 \times 10^{-18}$~km$^{-2}$yr$^{-1}$.  Taking into account the surface area of Umbriel (about $4 \times 10^6$~km), we get a formation rate of $10^{-11}$ per year, which gives a few percent probability of this impact happening over the age of the Solar System. Even more restrictive is the requirement that the impact had to happen within few hundred Myr after the establishment of the resonance, or else the inclination of Miranda would grow well beyond the observed $4.3^{\circ}$. Even before considering the probability of the correct impact geometry, we conclude that the probability of an impact sufficient to break the resonance in the required timeframe is well below one percent. We will therefore concentrate on the dynamical means of breaking the resonance.  

The dynamical excitation of all five orbits by the Ariel-Umbriel resonance is a direct consequence of the coupled dynamical nature of the Uranian satellite system. Due to the relative strength of satellite-satellite perturbations when compared to the perturbations from Uranus's oblateness, mean-motion resonances between Uranian satellites are overwhelmingly chaotic \citep{der88}. Furthermore, the non-unique correspondence between satellites and secular modes  \citep{las87, mal89} means that the eccentricities and inclinations of moons other than the resonant pair can be affected by the resonance. A good illustration of the mixing of secular modes is the isolated resonance capture at the beginning of the resonant encounter, best seen in Figs. \ref{damp} and \ref{damp2}. This sub-resonance is relatively well separated from the rest of the Ariel-Umbriel 5:3 multiplet, due to the fast precession rate of Miranda. For a brief period only the inclinations of Ariel and Miranda grow, indicating a resonant argument $5 \lambda_U -3 \lambda_A - \Omega_A -\Omega_M$, or more properly $5 \lambda_U -3 \lambda_A - \Omega_2 -\Omega_1$, where the numbers refer to secular modes. While the secular mode I1 is associated primarily with Miranda, dynamical coupling between Ariel and Miranda appears strong enough to make I1 a participant in a stably librating Ariel-Umbriel mean-motion resonant argument. The second and third secular modes (in both eccentricity and inclination) are generally associated with Ariel and Umbriel, respectively, but there is significant coupling between neighboring moons; fourth and fifth secular modes in $e$ and $i$ are approximately equaly shared between Titania and Oberon, with no one-to-one moon-mode correspondence \citep{las87}.

\begin{figure*}
\epsscale{.6}
\plotone{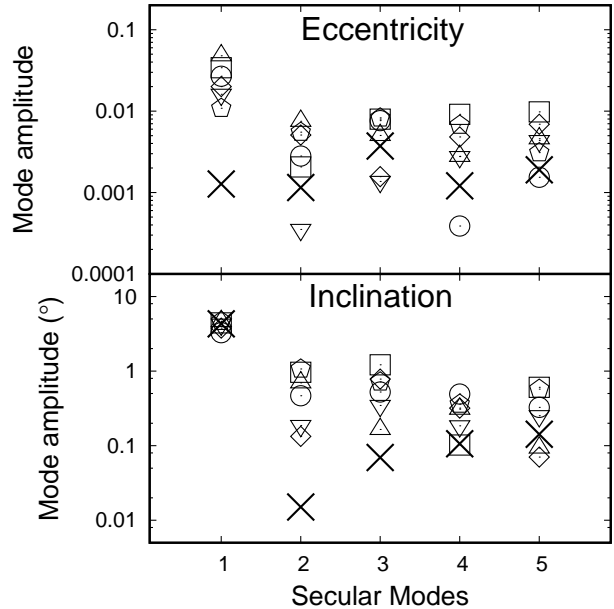}
\caption{The amplitudes of the five eccentricity and inclination secular modes of the Uranian system at the end of six different numerical simulations of Ariel-Umbriel 5:3 spin orbit resonance. The Xs plot the current strength of the secular modes, while the open geometrical shapes plot the simulation end-states (see Table \ref{table2} for details of each simulation). Here and elsewhere in the paper, the secular modes were determined for each orbital configuration by running $10^4$~yr simulations and applying frequency analysis to the output.\label{modes}}
\end{figure*}

\begin{table}[!ht]
\begin{center}
\caption{Details of the six simulations plotted in Fig. \ref{modes}, including the key to the symbols used to plot the end-states. Apart from the tidal properties of Uranus, Miranda and Ariel, the table also lists whether we used the Ariel and Umbriel masses from Table \ref{table1}, or updated values from \citet{jac14} (that have Ariel less massive than Umbriel). We also list secular resonances that happened between the breaking of 5:3 MMR and the end of the simulations. Simulation \#6 consisted of 5~Myr of highly accelerated evolution, followed by a longer simulation at a more realistic rate.\label{table2}}
\begin{tabular}{|l|c|c|c|c|c|c|c|c|}
\hline\hline
\# & Symbol & Duration & $(Q/k_2)_U$ & $(Q/k_2)_M$ & $(Q/k_2)_A$ & Masses & Secular Res. & Note\\
\hline\hline
1 & circle & 50 Myr & $4 \times 10^4$ & $10^6$ & $10^5$ & Table \ref{table1} & Modes 2-3 & Figs. \ref{accel_a}, \ref{accel_ei}\\
\hline
2 & square & 200 Myr & $1.33 \times 10^5$ & $10^5$ & $10^4$ & Table \ref{table1} &  Modes 3-4 & Fig. \ref{long}\\
\hline
3 & delta & 50 Myr & $4 \times 10^4$ & $10^4$ & $10^3$ & \citet{jac14} & none &  \\
\hline
4 & nabla & 50 Myr & $4 \times 10^4$ & $10^5$ & $10^4$ & \citet{jac14} & Modes 3-4? & \\
\hline
5 & diamond & 50 Myr & $4 \times 10^4$ & $10^6$ & $10^5$ & \citet{jac14} & none & \\
\hline
6 & pentagon & 5 Myr + 86 Myr & $4 \times 10^3$, $2 \times 10^5$ & $10^3$, $10^5$ & $10^2$, $10^5$ & Table \ref{table1} & several minor & two parts\\
\hline
\end{tabular}
\end{center}
\end{table}

Fig. \ref{modes} shows the amplitudes of each of the five secular modes of the Uranian satellites (in both eccentricity and inclination) at the end of six dissimilar numerical simulations of Ariel-Umbriel 5:3 MMR. These simulations all had similar initial orbital conditions but used different tidal parameters for the planet and the moons, with some runs also using slightly different masses for Ariel and Umbriel (see Table \ref{table2} for details of each simulation). We suspect that much of the variation in our results is purely stochastic and not due to different initial conditions, but many more (computationally expensive) integrations would be needed to confirm this. In our simulations the eccentricity modes are generally over-excited compared to the present state, but since they can be subsequently damped by eccentricity tides, here we will concentrate on the inclination modes which are more constant over time. We note that the inclination of Miranda (i.e. secular mode I1) is generally well-matched at the end of our simulations, obviating the need for the system to pass through the Miranda-Umbriel 3:1 MMR \citep{tw2, mal90}. However, the amplitudes of the other inclination modes (mode I2 in particular) are usually well above the observed values, seemingly ruling out Ariel-Umbriel 5:3 MMR crossing from happening. Here we face a conundrum: in order to obtain the large inclination of Miranda, some dynamical excitation event had to happen in the system's history, and Ariel-Umbriel 5:3 MMR is by far most recent commensurability for any tidal model. However, inclinations of the moons other than Miranda that were generated in the Ariel-Umbriel 5:3 MMR passage appear inconsistent with the present system. We find that the resolution of this discrepancy lies in the dynamical evolution of the moons following the mean-motion resonance crossing, which is discussed in the following sections.

\section{Secular Resonances} \label{sec:secular}

Uranian regular moons are similar in size to Saturn's mid-sized icy moons, but their secular dynamics is different, as the Uranians' orbital precession is not determined only by perturbations from the planet's oblateness, but also by significant moon-moon interactions \citep{las87, der88}. Moon-moon perturbations are mostly secular, but also have a component that depends on the proximity to the mean-motion resonances \citep{mal89}. Near-resonant perturbations make the secular frequencies vary considerably near MMRs, raising a possibility of secular resonances very close to mean-motion ones. Previously, \citet{cuk16} have found that Saturnian moons Tethys and Dione passed through a secular resonance immediately following the crossing of the Dione-Rhea 5:3 MMR. Unlike the secular modes of the Solar System \citep{md99}, fundamental precession frequencies of the Saturnian satellites are well-separated, even near mean-motion resonances. However, arguments of the type $\varpi+\Omega$ (i.e. the sum of the longitude of the pericenter and longitude of the ascending node) for each moon are changing very slowly, and can encounter resonances between each other near MMRs \citep{cuk16}. A clear signature of a secular resonance of this type are eccentricity and inclination of a moon changing in unison, with the $e$ and $i$ of one moon involved in the resonance being anti-correlated with those of the other. In other words, in a secular resonance of this type, one moon's orbit should get both more eccentric and inclined, while the other moon's orbit should become less eccentric and inclined at the same time. 

A clear example of a secular resonance is seen in Fig. \ref{accel_ei}, where the $e$ and $i$ of Ariel and Umbriel show major changes around 40~Myr, well after the breaking of their mutual MMR. The relevant resonant argument of $\varpi_2+\Omega_2-\varpi_3-\Omega_3$ (where the integer subscripts refer to secular modes, not the moons) can be deduced from D'Alembert's rules \citep{md99}; we confirm this by finding that its proxy $\varpi_A+\Omega_A-\varpi_U-\Omega_U$ (where letter subscripts refer to satellites) is librating around $180^{\circ}$ with a large amplitude between 37 and 41~Myr in the simulation. Similarly, Fig. \ref{long} features a secular resonance between modes 3 and 4 (i.e. with the argument $\varpi_3+\Omega_3-\varpi_4-\Omega_4$) close to the end of the simulation. We see some kind of secular resonance of this type in most of our simulations that went beyond the Ariel-Umbriel 5:3 MMR breaking, and all moons except Miranda being affected by at least one of these secular resonances. 

We also identify a simple $\varpi_3-\varpi_4$ (i.e. E3-E4 mode) secular resonance that can only occur before the Ariel-Umbriel 5:3 MMR (visible as a decrease in $e_U$ around 5~Myr in Fig. \ref{accel_ei}). This resonance makes the eccentricity of Umbriel drop while increasing $e$ of Titania and Oberon; it is caused by the retrograde adjustment to the precession of $\varpi_3$ just before the MMR due to Ariel-Umbriel near-resonant terms, making the precession of $\varpi_3$ as slow as that of $\varpi_4$. This resonance is unlikely to have been important in the real system, as it is likely that the moons' eccentricities would have been largely damped at that time.   
\begin{figure*}
\epsscale{.6}
\plotone{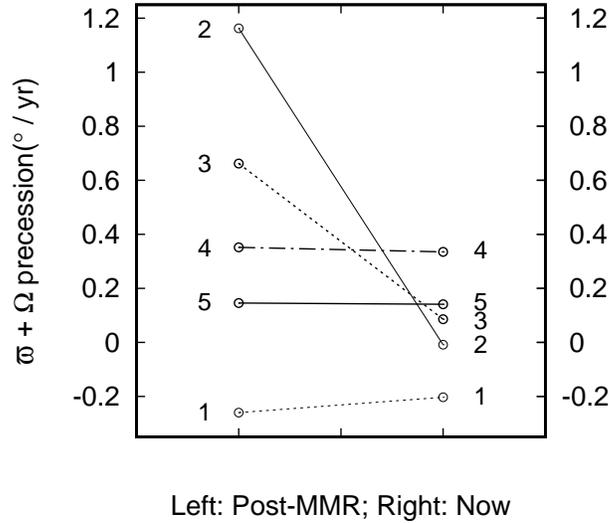}
\caption{Difference between the precession rate of arguments $\dot{\varpi}_{N} + \dot{\Omega}_{N}$ (where $1 \leq N \leq 5$) for each of the five Uranian system secular modes at the end of simulation \#5 (cf. Table \ref{table2}) and today. Precession rates for these arguments are significantly higher post-MMR in the case of modes 2 and 3 due to near-resonant perturbations on $\dot{\varpi}$. It is clear that the system must encounter secular resonances as it evolves from a near-resonant configuration to its present state.\label{periods1}}
\end{figure*}

While untargeted secular resonances appear to be common in our simulations, it would be valuable to have a more systematic picture of secular resonance crossings. Simulations that happened to end almost immediately after the mean-motion resonance breaking are particularly useful cases for understanding shifts in secular frequencies. Using the state vectors at the end of simulation \#5 in Table \ref{table2}, we produced Figure \ref{periods1} which compares the frequency relevant secular argument $\varpi+\Omega$ for each secular mode at the end of the simulation to the present Uranian system \citep[according to ][]{jac14}. Figure \ref{periods1} demonstrates that the precession of secular argument $\varpi+\Omega$ for modes 2 and 3 is expected to slow down significantly as Ariel and Umbriel move away from their resonance, while the modes 1, 4 and 5 are hardly affected. In the process, it is inevitable that the modes 2 and 3 would each go through secular resonances with modes 4 and 5, and also with each other. The precession rate of the argument $\varpi+\Omega$ for mode 1 (closely associated with Miranda) is negative and outside the range swept by other secular frequencies,\footnote{Using expressions derived by \citet{cuk18} it can be shown that the retrograde precession of the argument $\varpi_1+\Omega_1$ arises from Hamiltonian terms involving Uranus's oblateness and Miranda's inclination; prograde precession of equivalent terms for the other secular modes arises mostly from near-resonant moon-moon perturbations \citep{mal90}.} so Miranda is unlikely to be affected by any secular resonances of this type.

\begin{figure*}
\epsscale{.6}
\plotone{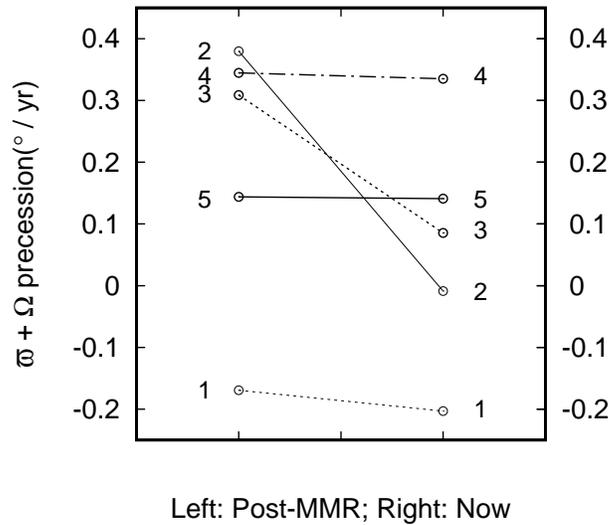}
\caption{Same as Fig. \ref{periods1}, but for the simulation \#4 in Table \ref{table2}. In this simulation the moons have evolved farther from the resonance than in Fig. \ref{periods1}, and the system appears to be close to a resonance between secular modes 2 and 4.\label{periods2}}
\end{figure*}

Since the secular resonances discussed here involve commensurabilities between relatively slow-changing arguments, it is desirable to study them using non-accelerated tidal evolution simulations. We find that full numerical simulations of secular resonances themselves are generally manageable, but that the tidal evolution between the resonances takes excessive time and arguably does not require full numerical integration. In Fig. \ref{periods1} it is obvious that the system is not very close to any secular resonances at the "post-MMR" stage. Therefore we turn to another integration featuring the Ariel-Umbriel 5:3 MMR crossing, simulation \#4 in Table \ref{table2}. Figure \ref{periods2} shows the precession rates of argument $\varpi+\Omega$ for each secular mode at the end of simulation \#4 and compares them to the present system.    

\begin{figure*}
\plotone{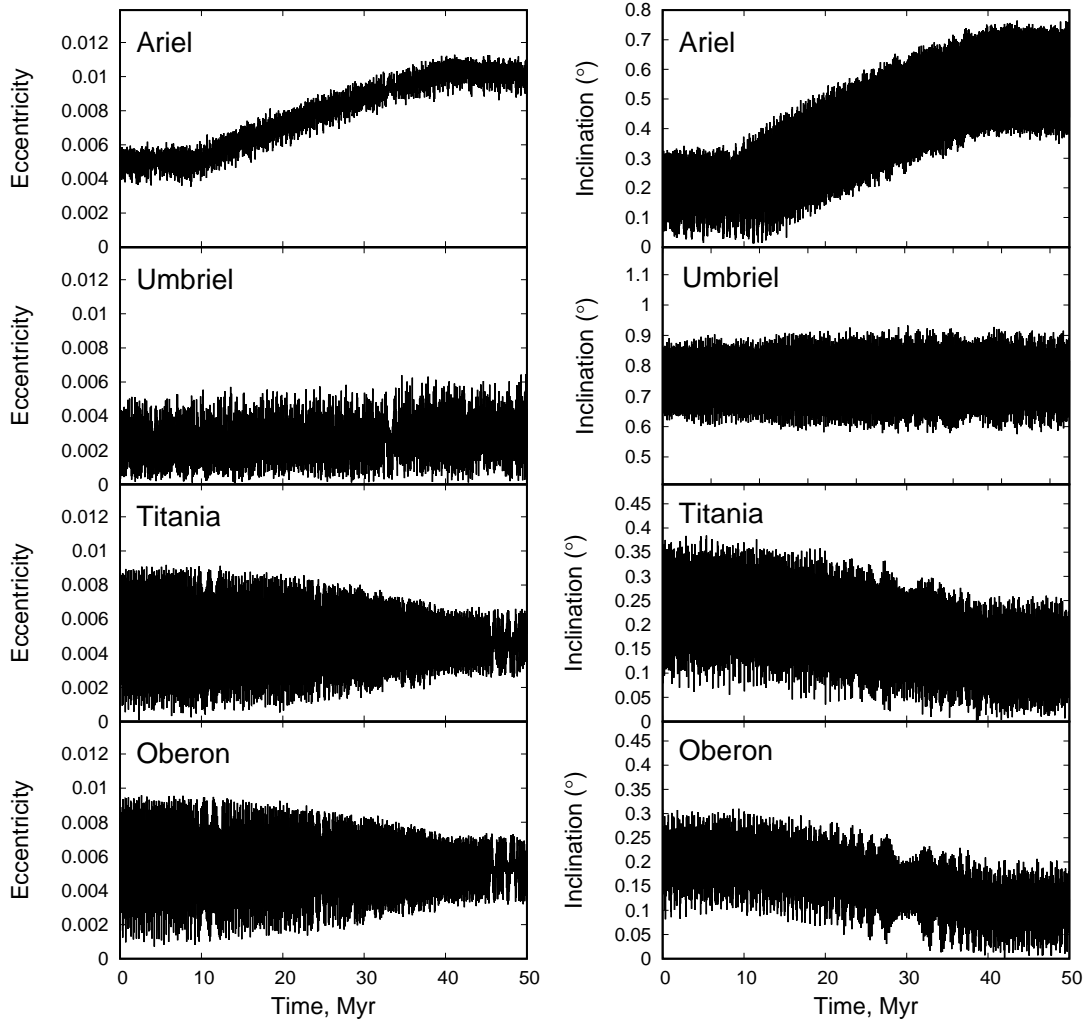}
\caption{Simulation of the secular resonance between modes 2 and 4, with the resonant frequency $\dot{\varpi}_2+\dot{\Omega}_2-\dot{\varpi}_4-\dot{\Omega}_4=0$. This simulation continues Ariel-Umbriel 5:3 MMR simulation \#4 from Table \ref{table2}, and uses our nominal $Q/k_2=4 \times 10^5$ for Uranus.\label{sec1}}
\end{figure*}

\begin{figure*}
\epsscale{.6}
\plotone{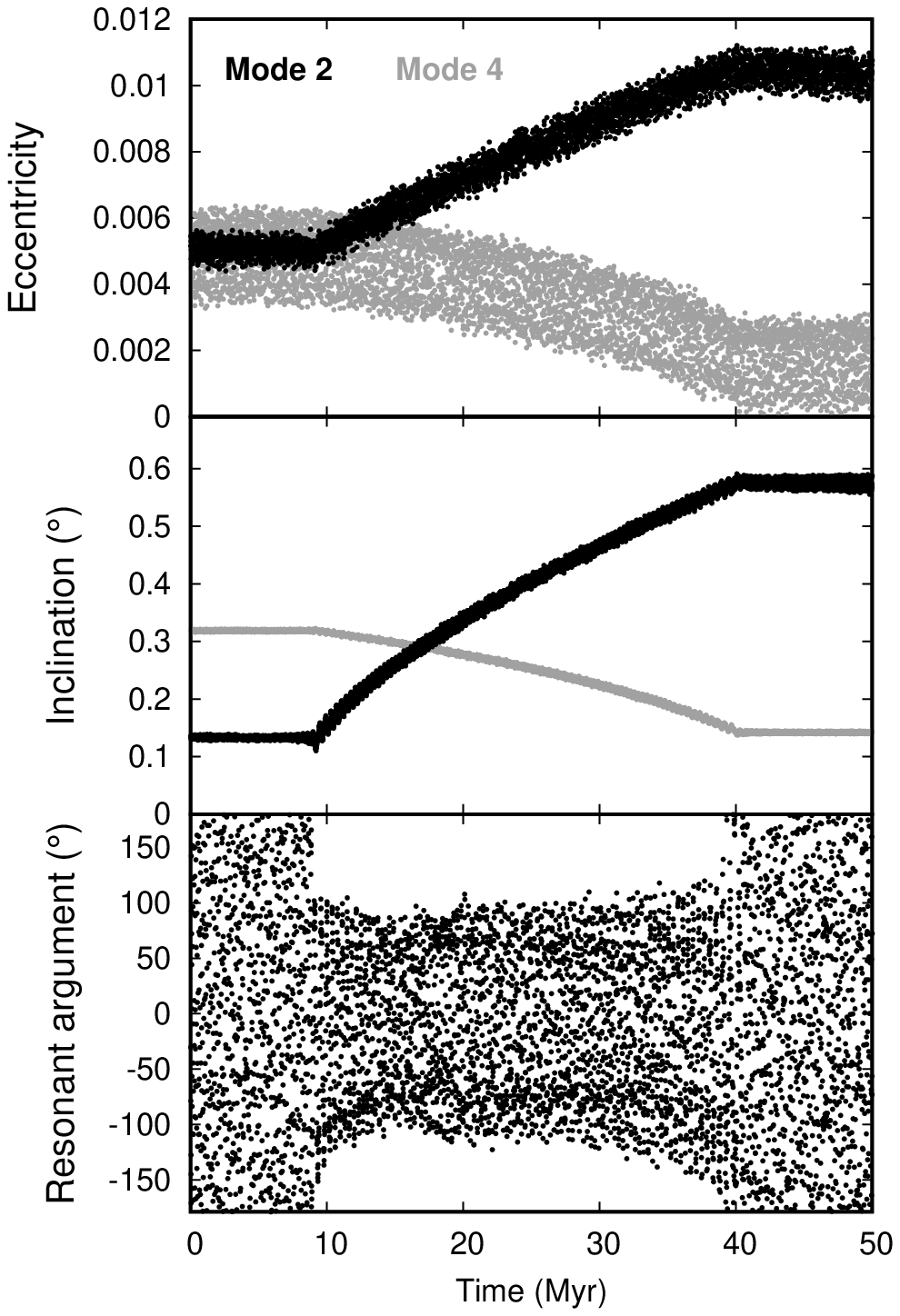}
\caption{The same simulation as shown in Fig \ref{sec1}, plotted in terms of secular modes rather than real satellites. The top panel plots eccentricities and the middle panel shows inclnations, with modes E2 and I2 plotted in black and modes E4 and I4 in gray. The bottom panel shows the resonant argument ${\varpi}_2+{\Omega}_2-{\varpi}_4-{\Omega}_4$. The normal mode conversion matrix was calculated from the system's configuration at 20~Myr into the simulation.\label{sec1mode}}
\end{figure*}

Figure \ref{periods2} suggests that the Modes 2 and 4 are close to a secular resonance. We therefore integrated the end state of simulation \#4 for 50 Myr. We used $Q/k_2=4 \times 10^5$ for Uranus, $Q/k_2=10^6$ for Miranda, and $Q/k_2=10^5$ for the other four moons; these tidal parameters may underestimate satellite tides, but enable us to observe the secular resonance isolated from other dynamical effects. The results of this simulation are plotted in Figs. \ref{sec1} and \ref{sec1mode}

As can be seen in Fig. \ref{sec1}, secular resonance between modes 2 and 4 increases $e$ and $i$ of Ariel, while decreasing the inclinations of Titania and Oberon and shrinking the amplitude of the variation in eccentricity of those two moons. This dissimilar behavior of $e$ and $i$ of both Titania and Oberon in this simulation is a consequence of the two moons' inclinations being dominated by secular mode I4, while their eccentricity is dominated by the secular mode E5 \citep[see ][ for the normal mode component table]{las87}. So while the amplitude of E4 decreases over time (Fig. \ref{sec1mode}), the average eccentricities of Titania and Oberon do not change much, as they are determined by the amplitude of the mode E5. The non-angular orbital elements of Umbriel appear to be largely unaffected, as are those of Miranda (not plotted). 

\begin{figure*}
\plotone{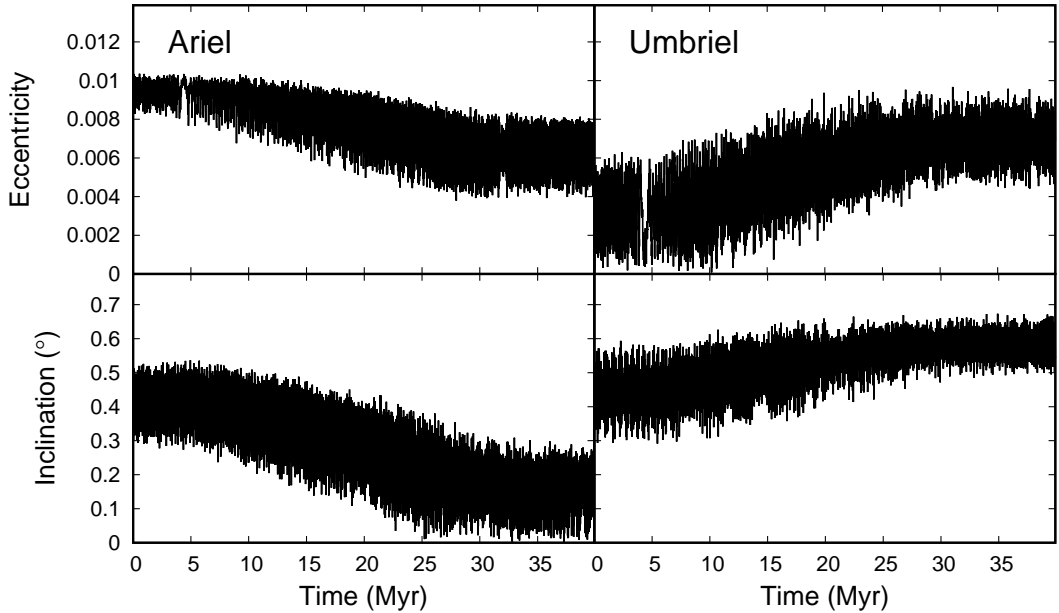}
\caption{Simulation of the secular resonance between secular modes 2 and 3, with the resonant frequency $\dot{\varpi}_2+\dot{\Omega}_2-\dot{\varpi}_3-\dot{\Omega}_3=0$. The initial conditions were chosen so that the system would go through the resonance immediately. We used our nominal $Q/k_2=4 \times 10^5$ for Uranus.\label{sec2}}
\end{figure*}

\begin{figure*}
\epsscale{.6}
\plotone{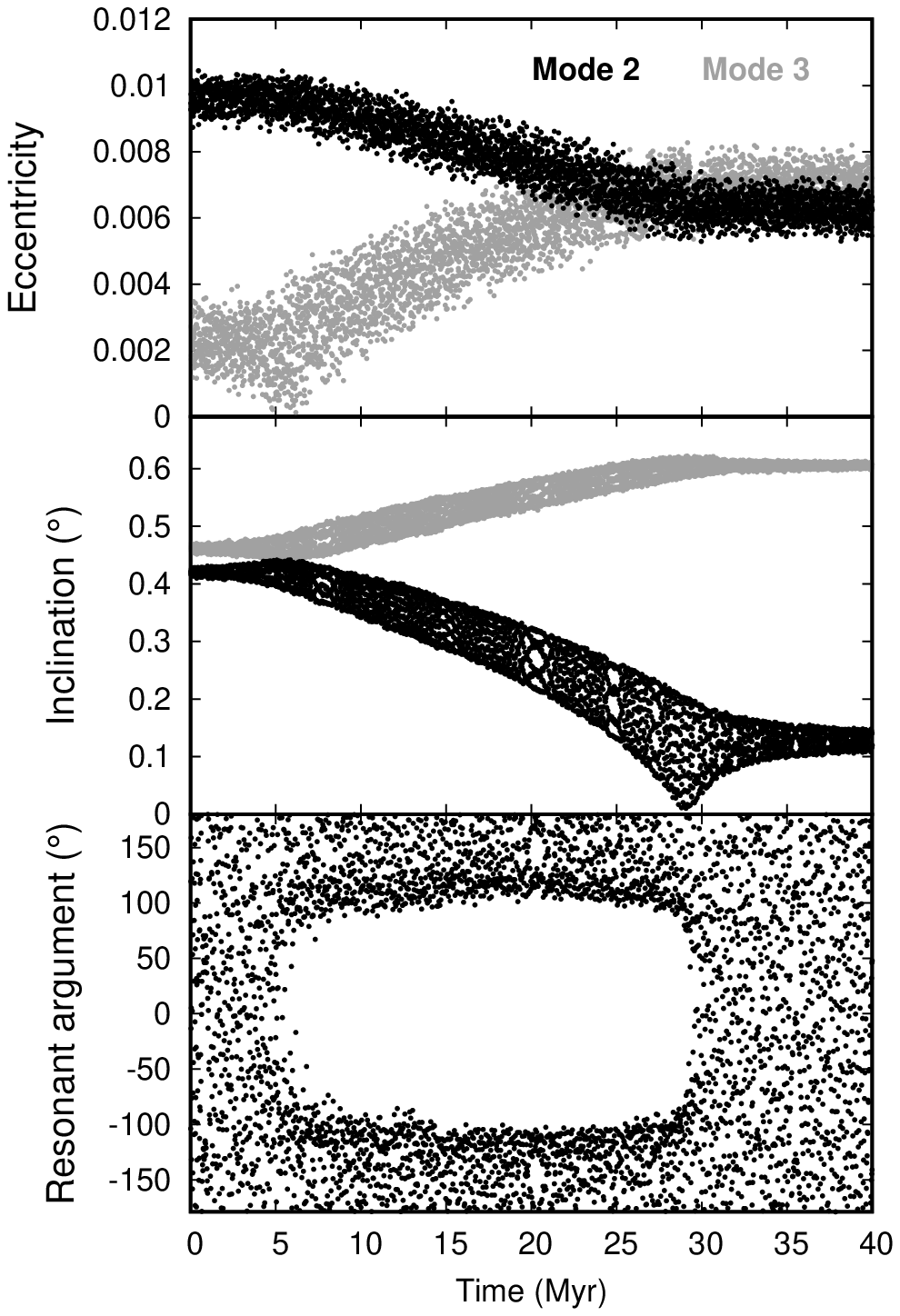}
\caption{The same simulation as shown in Fig \ref{sec2}, plotted in terms of secular modes rather than real satellites. The top panel plots eccentricities and the middle panel shows inclnations, with modes E2 and I2 plotted in black and modes E3 and I3 in gray. The bottom panel shows the resonant argument ${\varpi}_2+{\Omega}_2-{\varpi}_3-{\Omega}_3$. The normal mode conversion matrix was calculated from the system's configuration at 10~Myr into the simulation.\label{sec2mode}}
\end{figure*}

Both Figs. \ref{periods1} and \ref{periods2} imply that secular modes 2 and 3 should also experience a mutual secular resonance at some point following the MMR. As secular modes 2 and 3 (both in eccentricity and inclination) are strongly correlated with the orbits of Ariel and Umbriel, respectively, it is easy to detect this secular resonance through its proxy $\varpi_A+\Omega_A-\varpi_U-\Omega_U$. We were therefore able to place Ariel and Umbriel just interior to this secular resonance and allow the moons to enter it. The results of this numerical experiment is shown in Figs. \ref{sec2} and \ref{sec2mode}

Fig. \ref{sec2} shows the secular resonance between modes 2 and 3, in which the eccentricity and inclination of Ariel decrease while the $e$ and $i$ of Umbriel increase. At the end of the simulation the inclination of Ariel is relatively low, below 0.3$^{\circ}$. Fig. \ref{sec2mode} plots the resonance in terms of secular modes, and shows that the largest contribution to Ariel's final inclnation comes from mode I3, as mode I2 then has the amplitude of only about 0.1$^{^\circ}$. However, this is still about twice larger than the current amplitude of I2 found by \citet{las87}, and almost an order of magnitude larger than the value in \citet{jac14}, suggesting that more work is needed to establish whether secular resonance alone may not offer a full explanation for the present low amplitude of the I2 mode.  

We note that for secular resonance to evolve to very low amplitudes of I2, the amplitude of E2 must have been originally higher than that of I2 (in radians), or the secular resonance would break early because Ariel's eccentricity stopped being dominated by E2 (cf. Fig. \ref{sec2mode}). Larger amplitude for E2 over I2 is a minority outcome in our simulations of Ariel-Umbriel 5:3 MMR, but so far we have too few simulations and too many unexplored aspects of the problem to be able to determine probability of this outcome. 

The above examples demonstrate that secular resonances can re-distribute eccentricity and inclination among the moons of Uranus in the aftermath of the Ariel-Umbriel 5:3 MMR. The sole exception is Miranda, which is not affected by any of these resonances. While we have limited number of direct numerical integrations, simulations shown in Figs. \ref{sec1} and \ref{sec2}, combined with the timeline of secular resonances suggested by Figs. \ref{periods1} and \ref{periods2}, allow us to understand the post-MMR dynamical evolution of the Uranian system. It appears that secular resonances can heavily deplete the secular modes 2 and 4, while exciting the secular mode 3. We have simulated secular resonances of modes 2 and 3 with mode 5, and found they lead to small kicks in $e$ and $i$, rather than resonance capture, regardless of tidal evolution rates used. Fig. \ref{sec1} shows that the low inclinations of Titania and Oberon can be a result of secular resonance (such as between modes 2 and 4). Similarly, Fig. \ref{sec2} shows that the low inclination of Ariel may also largely be a result of a later secular resonance (such as between modes 2 and 3). Furthermore, it is likely that the eccentricities of Miranda and Ariel have been damped in the time since the Ariel-Umbriel 5:3 MMR (more in Section \ref{sec:dis}). The eccentricity of Umbriel is currently relatively high ($e_U=0.004$), and is consistent with Umbriel gaining $e$ in a secular resonance with Ariel. The largest remaining discrepancy between our simulations of the Uranian moons' dynamical history and their current orbits is that we predict the inclination mode I3 to be excited (to an order of a degree) while the present amplitude of I3 is only about $0.07^{\circ}$. We consider potential solutions in the next section.
 
\section{Spin-Orbit Resonances} \label{sec:spin}

In the Solar System, there are multiple examples of secular spin-orbit resonances, in which the precession period of a body's spin axis is commensurable with an orbital precession period in the system. Sometimes spin-orbit resonances lead to a chaotically varying obliquity for a planet, as is the case for Mars \citep{war73, tou93}, while at other times the spin axis can get captured into a stable resonance with orbital precession. Some of the examples of spin-orbit resonance capture in the Solar System include excitation of Saturn's obliquity by the $f8$ secular mode of the Solar System \citep{war04, ham04}, and the alignment of Koronis asteroid family spins due to a resonance with their nodal precession \citep{vok03}. More recently, spin-orbit resonances have been proposed as an important feature of compact multiple exoplanet systems \citep{mil19}. In each case, a change in one of the precession frequencies is required to achieve capture into the resonance, which can be supplied by planetary migration (in the case of Saturn), YORP radiative spin-down torque (for asteroids), or tidal spindown (for close-in exoplanets). 

In case of planetary satellites, no spin-orbit resonances of this type have been known so far. Most planetary satellites have tidally evolved spins, and currently rotate at the synchronous rate while their obliquities are purely forced and determined by the so-called Cassini States \citep{col66, pea69}. In Cassini States, as the moon's orbital plane precesses (by definition) around the Laplace plane, the satellite has a forced obliquity that causes its spin axis precess (relative to the Laplace plane) with the same period as the orbit. While the Cassini States have some ingredients of a spin-orbit resonance, they are usually not considered to be one. However, extensive past tidal evolution of the Moon placed it (billions of years ago) at a distance from Earth at which the spin and orbital precession frequencies were equal, leading to the switch between two Cassini States. The Moon must have had an extremely high forced inclination during and after this transition\citep{war75, cuk16b}. 

The Uranian system has some similarities with both other satellite and planetary systems. Like all close-in satellites, the five large Uranian moons are in synchronous rotation due to tides raised on them by the planet. Unlike the moons of Saturn, each Uranian moon's orbital precession involves several secular modes with different frequencies \citep{las87}, reminiscent of the planetary system \citep{md99}, offering numerous candidate frequencies for spin-orbit resonances. The reason spin-orbit resonances are relevant for the system's orbital history is that the obliquity tides can damp inclination, as long as the forced obliquity is sufficiently high \citep{chy89, che16}. Therefore, spin-orbit resonances have the potential to reconcile the discrepancy between the excited inclinations suggested by our simulations and the observed low inclinations of the four largest Uranian moons.

Unlike their rates of orbital precession, which are well determined, the axial precession rates of the Uranian moons have not been directly observed. Theoretical predictions of axial precession rates depend on the shapes and gravity fields of the satellites, which are for Miranda and Ariel based solely on limb coordinates, and are completely unconstrained for the other moons \citep{tho88}. \citet{che14} have previously calculated the expected moments of inertia of Uranian moons, under the assumption that they have uniform density and are in long-term hydrostatic equilibrium.  

\begin{table}[!ht]
\begin{center}
\caption{Rotation-related parameters of the major moons of Uranus, taken from Table 2 in \citet{che14} (except for the last column, which we calculated). The data on radii and orbits is from \citet{tho88} and \citet{jac92}, respectively. The gravity moments $J_2$ were calculated by \citet{che14} (we corrected an obvious typo in $J_2$ for Oberon) on the assumption of the hydrostatic equilibrium, and were used by us to calculate the moons' axial precession periods (see Eq. \ref{prec} and associated text).\label{chen}}
\begin{tabular}{|l|c|c|c|c|c|}
\hline\hline
Moon &  Radius & Density & Rotation & $J_2$ & Precesssion \\ 
\ & (km) & (kg m$^{-3}$) & period (d) & & period (yr)\\ 
\hline\hline
Miranda & 235.8 & 1200 & 1.412 & $6.10\times 10^{-3}$ & 0.1056 \\
\hline
Ariel & 578.9 & 1665 & 2.52 & $1.39\times 10^{-3}$ & 0.827 \\
\hline
Umbriel & 584.7 & 1399 & 4.13 & $6.13 \times 10^{-4}$ & 3.07\\
\hline
Titania & 788.9 & 1714 & 8.71 & $1.13\times 10^{-4}$ & 35.2\\
\hline
Oberon & 761.4 & 1630 & 13.47 & $4.93\times 10^{-5}$ & 124.7 \\
\hline
\end{tabular}
\end{center}
\end{table}
Table \ref{chen} lists the rotational parameters of the Uranian satellites, mostly based on data in Table 3 of \citet{che14}. We added a column with the calculated axial precession periods of the satellites, under the assumptions of uniform density, synchronous rotation, hydrostatic equilibrium, circular orbits, and and negligible obliquity for all the satellites. The precession rate was calculated using the expression \citep{war75}:
\begin{equation}
\alpha = {3 \over 2} n {(C-A) \over C} = 6 n J_2
\label{prec}
\end{equation}
where $A, B$, and $C$ are the smallest, intermediate, and largest moments of inertia of a moon, and $n$ its mean motion. For a hydrostatic satellite, $C-B=0.25(C-A)$ \citep[e.g. ][]{gar06}, therefore $(C-A)= (8/5) J_2 M R^2$, where $M$, $R$ and $J_2$ are the mass, radius and the dimensionless oblateness moment. For near-spherical uniform bodies $C = (2/5) M R^2$, giving us the final form of Eq. \ref{prec}. 

Comparison with secular frequencies of the system \citep{las87, mal90, jac14} suggests that Oberon's axial precession period (125~yr) is close to the secular mode I3 ($\simeq$130~yr). For the viability of resonant capture, it is crucial that Oberon's axial precession in the zero-obliquity, zero-libration case is slightly faster than the potentially resonant orbital perturbation, as obliquity forced by the resonance can only slow down the axial precession. 

In order to test the plausibility of a resonance between Oberon's axial precession and the secular mode I3, we use the numerical integrator {\sc r-sistem} that was developed by \citet{cuk16b} for the Earth-Moon system. This integrator has all the capabilities of {\sc simpl}, while also integrating the rotational motion of a satellite using the Lie-Poisson algorithm of \citet{tou94}. More detailed description of {\sc r-sistem} is available in the Methods section of \citet{cuk16b}.

\begin{figure*}
\epsscale{.6}
\plotone{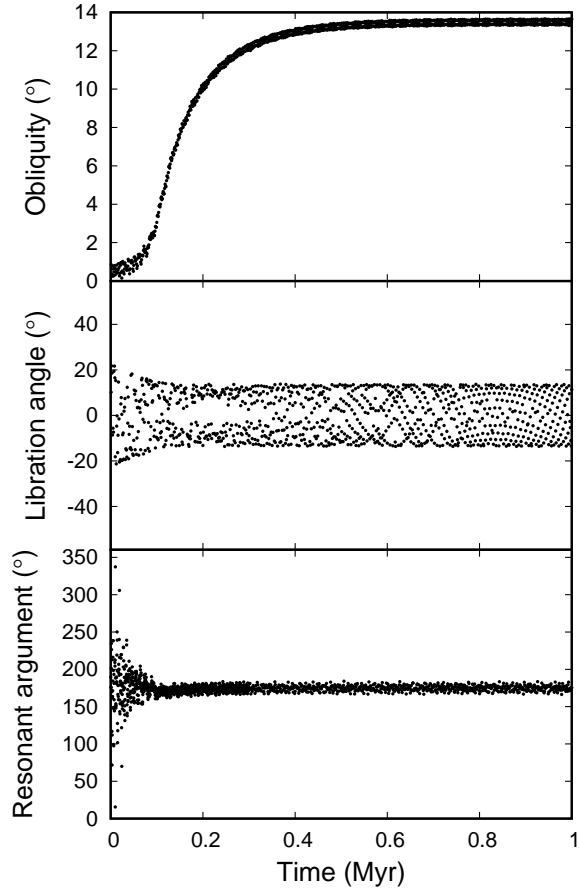}
\caption{Capture into secular spin-orbit resonance between the precession of the spin axis of Oberon and the nodal precession term associated with the secular mode I3. The inclination of Umbriel has been set to $i=0.6^{\circ}$, otherwise we use current orbits; Oberon is treated as a rigid triaxial body with moments dictated by hydrostatic equilibrium \citep{che14}. Initially, Oberon was placed in a synchronous rotation with $25^{\circ}$ longitudinal libration; as the rotational libration damps (middle panel) spin-orbit resonance capture occurs. Libration angle is not fully damped as we define it in three dimensions, so later in the simulation the libration angle reflects the obliquity rather than librations in longitude. The resonant angle is $\phi_O-\Omega_U$, where Umbriel's node has been used as a proxy for $\Omega_3$, the node of the secular mode I3.\label{oberon4}}
\end{figure*}

Figure \ref{oberon4} shows a numerical simulation of a spin-orbit resonance between Oberon and mode I3 done using {\sc r-sistem}. We used the nominal orbits for the current Uranian system (Table \ref{table1}) with one exception: we increased the inclination of Umbriel to $0.6^{\circ}$ (to demonstrate that it can be damped). For the rotation of Oberon, we assumed a synchronous rotation rate, but we created an in-plane misalignment between the orientation of the long axis of Oberon and the direction to the planet of about $25^{\circ}$. We used $Q/k_2=10^4$ for Oberon. The initial misalignment led to longitudinal librations that were damping relatively fast (middle panel), until the spin-orbit resonance was reached at about $10^5$~yr, and the obliquity of Oberon was excited (top panel). The resonant argument $\phi_O-\Omega_3$ (where $\phi_O$ is the longitude of the intersection of the equatorial planes of Oberon and Uranus) is plotted in the bottom panel, clearly demonstrating capture into resonance. Eventually Oberon settles into a stable resonant state with an obliquity of about 13$^{\circ}$.
    
\begin{figure*}
\epsscale{.55}
\plotone{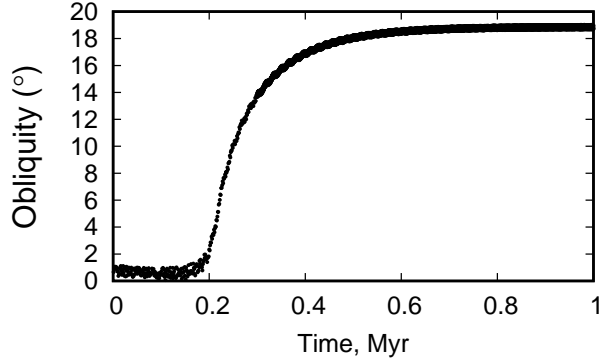}
\caption{Similar to the top panel in Fig. \ref{oberon4}, but now showing the Oberon-Umbriel secular spin-orbit resonance with satellite masses from \citet{jac14}, rather than those from Table \ref{table1}. The lower mass of Ariel decreases the precession rate of mode I3, necessitating a higher forced obliquity of Oberon for resonance capture.\label{oberon5}}
\end{figure*}

The simulation shown in Fig. \ref{oberon4} used Ariel and Umbriel masses listed in Table \ref{table1}. Since the spin-orbit resonance is sensitively dependent on the frequency of the I3 mode, we also performed a simulation with masses taken from \citet{jac14}. The smaller mass of Ariel in \citet{jac14} makes the frequency of I3 slightly lower, requiring a higher forced inclination for Oberon to maintain the resonance, and Fig. \ref{oberon5} shows that this new inclination is about 19$^{\circ}$. This higher forced obliquity also leads to an even faster inclination damping for secular mode I3 \citep{chy89}. We conclude that the spin-orbit resonance is likely to be robust against the uncertainty in satellite masses. 

\begin{figure*}
\epsscale{.6}
\plotone{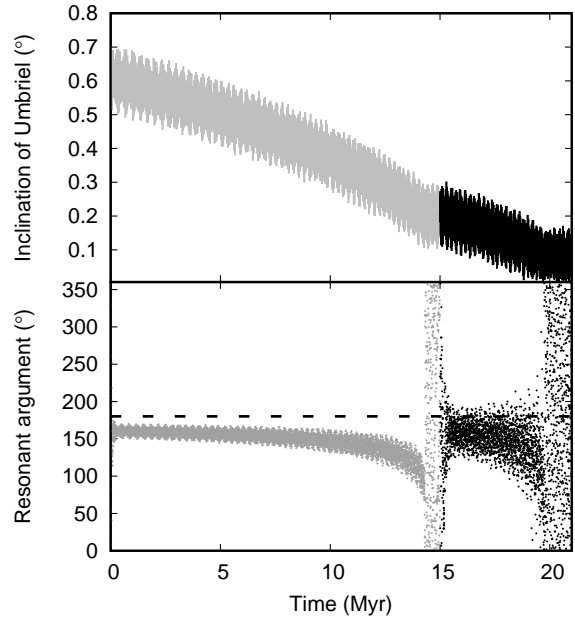}
\caption{Damping of Umbriel's inclination (top panel, as a proxy for the amplitude of the secular mode I3) through the secular spin-orbit resonance with Oberon's axial precession. The bottom panel shows the (proxy) resonant argument $\phi_O-\Omega_U$. The gray line plots a simulation that used $Q/k_2=3.3 \times 10^3$ for Oberon, which was the highest dissipation that allowed for resonance capture in our simulations. The first simulation was terminated once the resonance was broken at about 14.5 Myr, with the average final inclination $i_U=0.2^{\circ}$. We then ran another simulation continuing the orbital evolution (black line), but with $Q/k_2=10^4$ for Oberon. Lower dissipation allowed the spin-orbit resonance to be maintained at lower inclinations of Umbriel, with the final $i_U \simeq 0.07^{\circ}$, close to its present value.\label{oberon8}}
\end{figure*}

We have also run longer spin-orbit resonance simulations to demonstrate that obliquity tides in this spin-orbit resonance act to damp the amplitude of the I3 mode. Due to slower speed of {\sc r-sistem} relative to {\sc simpl}, we had to sacrifice some of the elements of the simulation for speed. We decided to remove Miranda from the simulation, as it does not directly affect the spin-orbit resonance of Oberon, and this modification enables us to use a longer timestep (as Miranda's orbital period was the shortest timescale in the problem). Also, we started the simulation shown in Fig. \ref{oberon8} with the highest tidal dissipation within Oberon that still allowed spin-orbit resonance capture ($Q/k_2=3.3 \times 10^3$), which was determined empirically.    

Figure \ref{oberon8} shows the evolution of Umbriel's inclination during a two-part four-moon simulation. In the first simulation described above (gray line in Fig. \ref{oberon8}), the inclination of Umbriel is declining monotonically, with an approximately constant $i(di/dt)$\citep{chy89}. The resonance breaks when the mode I3 (approximately equivalent to the mean inclination of Umbriel) reaches about $0.2^{\circ}$. The resonance breaks because both the offset of the resonant argument's libration center from $180^{\circ}$ and the argument's libration amplitude (Fig. \ref{oberon8}, bottom panel) are inversely proportional to the amplitude of the I3 mode (i.e. approximately Umbriel's mean inclination). As the strength of the I3 mode is damped, librations of the argument $\phi_{O}-\Omega_3$ grow and shift away from $180^{\circ}$, until they exit the stable island and the resonance is broken. This end-point of the resonance is not inevitable, but is dictated by our choice of Oberon's tidal properties. The center of librations in the resonance is offset from $\phi_{O}-\Omega_3=180^{\circ}$ because of the presence of dissipation, and the amount of offset is proportional to Oberon's $k_2/Q$ ratio. To explore how weaker tidal dissipation affects the resonance, we continued the simulation using $Q/k_2=10^4$ for Oberon, and once again introduced a free rotational (longitudinal) libration of Oberon in order for resonance capture to occur. Oberon re-enters the spin orbit resonance and the damping of Umbriel's inclination continues (at a slower pace) until it is again broken when the amplitude of the I3 mode is $i_3=0.07^{\circ}$ (black line), which is close to the current value.

How realistic is the capture into Oberon's secular spin-orbit resonance in the real system? There are two plausible routes to resonance capture. One is collisional, in which an impact induces rotational librations of Oberon (within synchronous state) that are large enough to enable resonance capture during libration damping. Using the approach of \citet{lis85}, we calculate that an impactor large enough to induce $20^{\circ}$ longitudinal librations of Oberon (once again assuming a hydrostatic shape, Table \ref{chen}) should create a D=70~km crater on Oberon. According to \citet{zah03}, craters of this size are formed on Oberon by cometary impacts 1-2 times in a Gyr. Therefore, establishment of Oberon's spin-orbit resonance after the Ariel-Umbriel 5:3 MMR (and associated secular resonances) excited the I3 mode through cometary impacts is plausible. Additionally, as Oberon is the outermost regular satellite of Uranus, it may experience collisions with the irregular satellites of Uranus \citep{gla98, gla00, kav04}, some of which may have long-term variations in eccentricity due to solar perturbations \citep{car02, nes03}. The rate of irregular satellite impacts on Oberon is currently unknown, but this process could in principle significantly increase the likelihood of secular spin-orbit resonance capture.

\begin{figure*}
\epsscale{.6}
\plotone{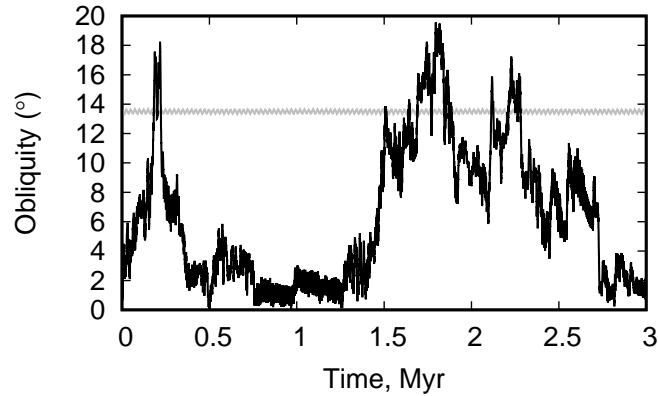}
\caption{Obliquity of Oberon (black line) during a simulation of Ariel-Umbriel 5:3 MMR. Gray line plots the resonant obliquity in the absence of MMR (cf. Fig.\ref{oberon4}). The mean-motion resonance clearly induces chaotic dynamics in the precession of secular mode I3, which in turn appears to make the obliquity of Oberon chaotic through the Oberon-I3 secular spin-orbit resonance.\label{oberon11}}
\end{figure*}

The other route to spin-orbit resonance capture is purely dynamical. The spin-orbit resonance shown in Figs. \ref{oberon4}--\ref{oberon8} is regular and devoid of chaotic behavior. However, we know that all of the secular modes of the Uranian system experience chaotic behavior during the Ariel-Umbriel 5:3 MMR (Figs. \ref{accel_ei}--\ref{long}). In order to explore the effects of chaos on the spin of Oberon, we took a simulation of MMR crossing (very similar to the one shown in Figs. \ref{accel_a}--\ref{accel_ei}) which still had Ariel and Umbriel in the resonance at 50~Myr, and continued it using {\sc r-sistem}, with the spin dynamics of Umbriel taken into account. Oberon was initially given small obliquity and rotational libration amplitude. The results are plotted in Fig. \ref{oberon11}: Oberon's inclination varies chaotically between about zero and $20^{\circ}$ (for the satellite masses used, the resonant obliquity outside the MMR is $13^{\circ}$, as in Fig. \ref{oberon4}). This is clearly a consequence of the mode I3 being affected by the MMR, which in turn couples the precession of Oberon to the Ariel-Umbriel resonance. 

So far we have not seen any exits from the Ariel-Umbriel 5:3 MMR in simulations that included Oberon's rotation. Slower speed of integrations using {\sc r-sistem} and unpredictability of the resonance breaking makes it impractical to numerically simulate this process with our current resources. We simply note that the obliquity of Umbriel is near the resonant value for a large part of the simulation shown in Fig. \ref{oberon11}. It is likely that for some of these values of obliquity and resonant argument Oberon will find itself in the stable spin-orbit resonance once the MMR is broken. Our simulations also indicate that, if Oberon's obliquity is higher than resonant after MMR breaking, capture into spin-orbit resonance through obliquity damping appears to be certain (for realistic tidal properties of Oberon). Future work should be able to establish the probability of Oberon entering the stable spin-orbit resonance due to the breaking of Ariel-Umbriel 5:3 MMR.

If the spin-orbit resonance is established at the exit from the MMR, the secular mode I3 could be gradually damped, and this damping can be continued during the I2-I3 mode secular resonance. This parallel evolution within both the secular and spin-orbit resonance is plausible, as both have natural timescales on the order of $10^8$~yr, but direct numerical integrations will be necessary to test this hypothesis. 


\section{Three-Body Resonances} \label{sec:3br}

While we can be confident that Ariel-Umbriel 5:3 MMR was the last major mean-motion commensurability between Uranian moons, our work has uncovered a number of unexpected dynamical features, including secular and spin-orbit resonances, that we did not foresee at the start of the project. It is natural to ask whether there are any additional resonances (or other dynamical events) that happened between the Ariel-Umbriel 5:3 MMR and the present, which should be taken into account when comparing the present system to our simulations. We decided on a brute-force approach, dividing up the total orbital evolution of Ariel (measured by the Ariel-Umbriel period ratio) between the MMR and the present into ten sections, and covering each of them by a simulation with $Q/k_2=2 \times 10^4$ for Uranus. These runs were therefore accelerated about 20 times over our nominal $Q/k_2=4 \times 10^5$ case, and ten 10 Myr simulations approximate about 2 Gyr of the tidal evolution we would have in the nominal case. 

\begin{figure*}
\epsscale{.6}
\plotone{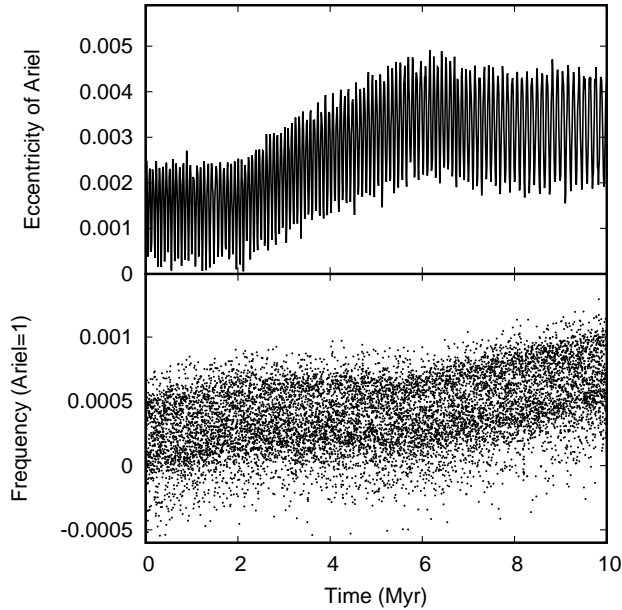}
\caption{Top: eccentricity of Ariel in a simulation of Ariel-Umbriel-Titania three-body resonance that should have happened more recently than Ariel-Umbriel 5:3 MMR. We used $Q/k_2=10^5$ for Uranus, four times above the nominal rate. There is a brief capture into the resonance, during which the eccentricity of Ariel is excited to $e_A \geq 0.003$. Bottom: evolution of the three-body resonant frequency expressed in terms of Ariel's mean motion: $4(a_U/a_A)^{-3/2}+2(a_T/a_A)^{-3/2}-3$. The offset from zero reflects the precession terms in the resonance.\label{fish4}}
\end{figure*}

In this sweep for more recent resonances, we found only one new dynamical feature of note. It happens when the Ariel-Umbriel 5:3 inequality has a period of about 86 days (the period of this inequality is very long just after the exit from 5:3 MMR, and about 62 days at the current epoch). At this location we see temporary captures of Ariel, Umbriel or both into an apparent eccentricity-type resonance, and we also see much smaller kicks to their inclinations. Figure \ref{fish4} plots the eccentricity of Ariel in a somewhat slower ($Q/k_2=10^5$ for Uranus) simulation centered around the resonance. We find that this resonance is a combination between two two-body near-resonances: Ariel-Umbriel 5:3 and Umbriel-Titania 2:1 inequalities. These two inequalities have opposite signs: while Ariel is past 5:3 MMR with Umbriel (so that their inequality is prograde), Umbriel is interior to the 2:1 MMR with Titania (which makes their near-resonant frequency retrograde). Therefore, this resonant argument's mean-motion part is $4 \lambda_U - 3 \lambda_A + 2 \lambda_T$, which make it a third order three-body resonance \citep{nes98, qui11, gal16}. For our purposes, this resonance is important as being likely the most recent process to increase the eccentricity of Ariel, and is therefore crucial for estimating Ariel's eccentricity damping timescale (Section \ref{sec:dis}).

\begin{figure*}
\epsscale{.6}
\plotone{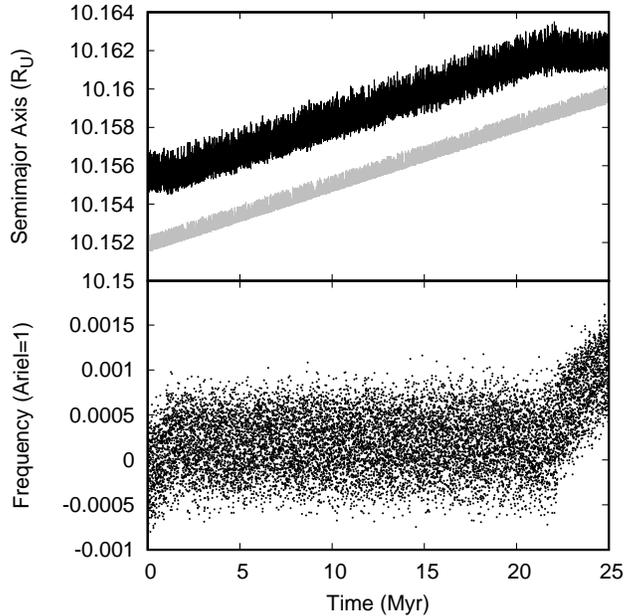}
\caption{Top: semimajor axis of Umbriel (black) in a simulation of a Ariel-Umbriel-Titania three-body resonance predating the Ariel-Umbriel 5:3 MMR. Here we used $Q/k_2=4 \times 10^4$ for Uranus, order of magnitude below our nominal value. Resonance capture is relatively long-lived, and the length of the capture depends on the dissipation within the moons, the eccentricities of which are all coupled. The resonance appears to be combination of $6 \lambda_U - 3 \lambda_A - 2 \lambda_T - \varpi_N$  arguments, where $N=\{2, 3, 4\}$, and all three arguments are observed to be chaotic. The gray line plots the semimajor axis equivalent to the orbital period that is 2.94 times that of Miranda. During the resonance, relative tidal evolution of Umbriel and Miranda are the same, effectively "stopping the clock" on the timing of their past resonances. Bottom: evolution of the three-body resonant frequency expressed in terms of Ariel's mean motion: $6(a_U/a_A)^{-3/2}-2(a_T/a_A)^{-3/2}-3$. The offset from zero reflects the precession terms in the resonance. \label{3br1}}
\end{figure*}

The detectability of the 3rd order three-body Ariel-Umbriel-Titania resonance postdating the Ariel-Umbriel 5:3 MMR raises the possibility of other three body resonances being prominent in the past of the system. While we did not have the time or the resources for a thorough search for new resonances at earlier epochs, there is an obvious candidate based on the above results. Before the Ariel-Umbriel 5:3 MMR the associated near-resonant frequency was retrograde, just like the Umbriel-Titania 2:1 inequality. Therefore the combination of these two near-resonances would have the mean-motion argument of $6 \lambda_U - 3\lambda_A - 2 \lambda_T$, and being of the first order it would be potentially much stronger than its more recent cousin. Figure \ref{3br1} shows an integration using Uranian $Q/k_2=4 \times 10^4$ through this resonance (assuming initially low-$e$ and low-$i$ orbits for all the moons). The first order three-body resonance is clearly stronger and the moons spend a much longer time in the resonance. However, since the resonance is also much wider at low eccentricity than its third-order relative, there is chaotic overlap between the three subresonances that are each related to the eccentricity of Ariel, Umbriel or Titania. This produces chaotic variations in eccentricity for all three moons. The resonance breaks when the eccentricity of Umbriel reaches about $e_U=0.008$, with the eccentricities of all five moons being of same order (Miranda appears to be affected by secular perturbations of Ariel, while Oberon shares Titania's secular modes).

In terms of capture and escape, this resonance has similarities to the Ariel-Umbriel 5:3 MMR, as it is chaotic and can be broken when the moons' eccentricities reach certain values. This implies that, if there is strong enough tidal dissipation within moons and eccentricity is kept in check, this resonance could in principle last a very long time. Unlike in the case of the second-order Ariel-Umbriel 5:3 MMR, this three-body resonance does not affect inclinations appreciably, so there is no limit from growing inclinations on how long the residence in resonance can last. Figure \ref{3br1} shows that while the three-body resonance is active, tidal expansion of Umbriel's orbit is enhanced and is very close to that of Miranda. Therefore, time spent in this three-body resonance is almost completely "invisible" to calculations of the timing of the past Miranda-Umbriel 3:1  MMR (cf. Figure \ref{peale}). This has direct implications for the constraints we may be able to place on the tidal properties of Uranus based on past MMR crossings.  

\section{Discussion} \label{sec:dis}

\subsection{Tidal Heating of Miranda} \label{subsec:miranda}

Miranda has dramatic surface features that indicate extensive geologic activity, some of which is relatively recent \citep{ple87}. According to crater counts, some of the most prominent surface features are large coronae. Two coronae, Arden and Inverness, appear to be only about 1 Gyr old, while others are significantly older, according to heliocentric impactor flux modeling \citep[][ see also Section \ref{subsec:uranus}]{zah03}. A recent study of the morphology of these features, coupled with modeling of Miranda's interior \citep{bed15} has found that heat flows of 30-100~mW~$m^{-2}$ are necessary to produce the Arden corona. Can the excitation of Miranda's eccentricity during the Ariel-Umbriel 5:3 resonance provide sufficient tidal heating for the formation of the young coronae?

In our simulations we mostly focused on the overall dynamics of the Ariel-Umbriel resonance, which has major effects on Miranda's orbit but is only weakly affected by Miranda itself. Therefore, we did not explore the phase space of Miranda's tidal proprieties in great detail. However, we did run some integrations in which the tidal response of Miranda was greatly enhanced (to simulate melting; similar to what was done in Figs. \ref{damp} and \ref{damp2} for Ariel) in order to see whether there are effects on the Ariel-Umbriel resonance. It turns out the resonance stays practically unaffected, but there are major consequences for Miranda.

\begin{figure*}
\plotone{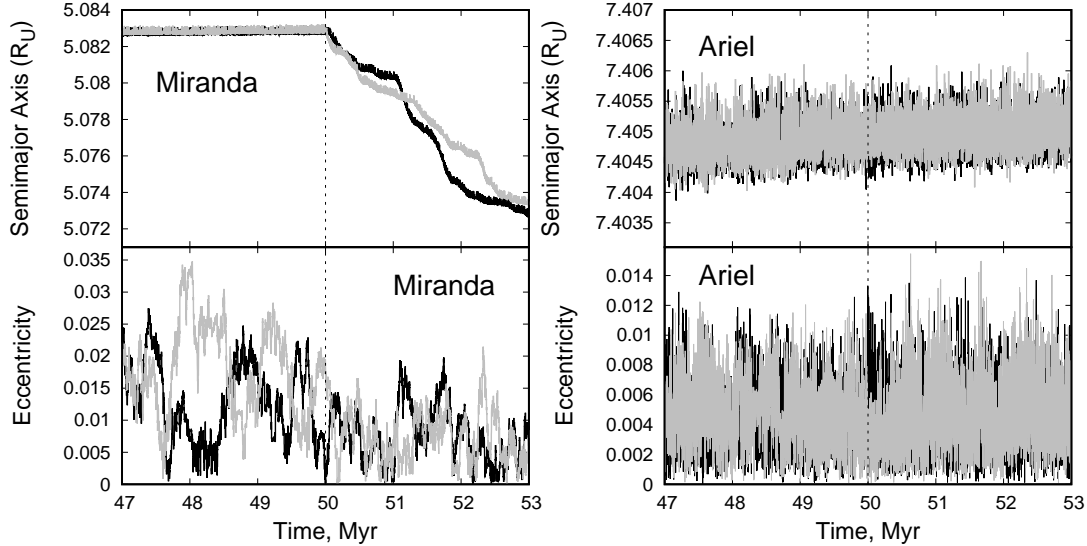}
\caption{Two simulations of the Ariel-Umbriel 5:3 MMR during which we greatly increased the rate of tidal dissipation within Miranda, as a proxy for internal melting. At the start of the simulations, we used $Q/k_2=2 \times 10^5$ for Uranus and $Q/k_2=10^5$ (black line) and $Q/k_2=10^6$ (gray line) for Miranda. At 50 Myr into the simulation we changed Miranda's tidal response to $Q/k_2=100$ in both simulations \citep[similar to that of Enceladus;][]{lai12}. Miranda's average eccentricity becomes somewhat lower ($e_M \leq 0.02$) but it remains chaotically excited, while the semimajor axis of Miranda shrinks due to strong eccentricity tides. There is no appreciable effect on Ariel and other moons, or Miranda's inclination, which behave the same as in other Ariel-Umbriel 5:3 MMR simulations. \label{miranda}}
\end{figure*}

Figure \ref{miranda} shows the orbital elements of Miranda and Ariel in two simulations of Ariel-Umbriel 5:3 MMR in which the tidal dissipation of Miranda was abruptly increased, to an Enceladus-like value. Chaotic interaction between Ariel and Miranda continues, even though the eccentricity of Miranda now does not rise above $e_M=0.02$. There is no visible change in the behavior of Ariel's orbit. However, Miranda's semimajor axis now drops relatively rapidly due to tidal dissipation within Miranda \citep[cf. ][]{kau64, efr14}. This semimajor axis drop has no effect on the excitation of Miranda's eccentricity and inclination by Ariel or other moons, which is driven by secular dynamics. This is a very important result, as it offers a mechanism of tidal heating that is not limited by the power supplied by the outward tidal evolution rate of the moons \citep{pet15}.

Figure \ref{miranda} suggests that the semimajor axis of Miranda can decrease by about 0.01~$R_U$ (0.2\%) over 3 Myr (about $10^{14}$~s). Since $(\dot{E}/E)=-(\dot{a}/a)$ and the orbital energy of Miranda is about $1.4 \times 10^{27}$~J, the power lost from the orbit through tidal dissipation is about $3 \times 10^{11}$~W, yielding an energy flux of 0.4~W~m$^{-2}$, which is well in excess of 31-112~mW~m$^{-2}$ required by \citet{bed15} for formation of Arden Corona. Therefore even with $Q/k_2=10^3$, Miranda would produce enough heat to explain the formation of the youngest coronae (note that Miranda's average eccentricity during Ariel-Umbriel MMR would likely be higher for larger $Q/k_2$, enhancing tidal power available). The tidal heating available from Miranda's orbital shrinking due to eccentricity tides, ultimately caused by chaotic secular perturbations from other moons, is two orders of magnitude higher than that available from Miranda's own tidal orbital expansion, even if an optimistic $Q=11,000$ is assumed for Uranus.  Therefore we conclude that the Ariel-Umbriel 5:3 MMR is almost certainly the cause of the most recent episode of tidal heating on Miranda. Given that the coronae on Miranda have a range of ages \citep{zah03}, it is likely that there were other dynamical events preceding the Ariel-Umbriel 5:3 MMR that have caused major tidal heating episodes for Miranda.

\subsection{Tidal Heating of Ariel} \label{subsec:ariel}

In contrast to Miranda's freedom to migrate inward due to tidal dissipation, Ariel is the main driver of the mean-motion resonance that is exciting the moons' orbits, and any inward movement of Ariel would break the resonance (as seen in Fig. \ref{damp}). Therefore, power available for the tidal heating of Ariel is limited to that available from tidal expansion of Ariel's orbit, which is on the order of 1 mW m$^{-2}$ \citep{pet15}, Therefore the Ariel-Umbriel 5:3 MMR is almost certainly not the root cause of geological activity on Ariel, which requires 28-92~mW~m$^{-2}$ \citep{pet15}. Lack of significant heating of Ariel is consistent with the dynamical requirement that Ariel's tidal response must be weak in order to allow Ariel and Umbriel to eventually exit their 5:3 MMR (Section \ref{sec:mmr}).

None of the terrains on Ariel appear to be as recent as the youngest features on Miranda \citep{zah03}, which further supports our results that the last episode of tidal heating was likely caused by Ariel-Umbriel 5:3 MMR, and that it had significant geophysical consequences on Miranda but not Ariel. At this point it is not clear what past dynamical mechanisms could have resulted in enough tidal heating to affect Ariel's surface, but our results for Miranda offer us some clues. Contrary to common expectation, it appears that secular perturbations, rather than mean-motion ones, may hold more promise in generating sufficient tidal heating. This is because secular resonances are not restricted to the power generated by orbital expansion due to tides, as mean-motion resonances are. The complexities of orbit-orbit and spin-orbit secular resonances we found in Sections \ref{sec:secular} and \ref{sec:spin} hint that more interesting dynamical behavior may be discovered once an even more ancient past of the Uranian system is studied using direct numerical integration.
 
\subsection{Tidal Response of Uranus} \label{subsec:uranus}

In this paper we presented theoretical reconstruction of part of the dynamical history of the Uranian system. We in particular focused on matching the current orbital inclinations of the moons, which are much harder to change without resonances, unlike eccentricities which are routinely damped by satellite tides. In this section we will address the absolute chronology of the resonances we studied and its implication for Uranus's tidal response, as well as what current eccentricities of the moons can tell us about their tidal responses.

\begin{table}[!ht]
\begin{center}
\caption{Tidal properties of the five large Uranian moons. The first column gives the moons' eccentricity damping constants, which need to be multiplied by $Q/k_2$ to obtain the relevant eccentricity damping timescales. The next two columns list the moons tidal Love numbers $k_2$ as estimated respectively for icy and rocky bodies, and are calculated using the specified rigidity $\mu$. The last column shows the moons' relative orbital expansion in 1~Gyr, assuming our nominal $Q_p=40,000$. The moons' parameters were taken from Table \ref{table1}, and we used expressions given in \citet[][ Section 4.10]{md99} to calculate the derived quantities.\label{tidal}}
\begin{tabular}{|l|c|c|c|c|}
\hline\hline
Moon &  $\tau_e (k_2 / Q)$ &``Ice'' $k_2$ &``Rock'' $k_2$ & $(\dot{a} / a) (40,000 / Q_p)$ \\ 
\ & [yr] & ($\mu=4 \times 10^9$~Pa) & ($\mu=5 \times 10^{10}$~Pa)& [Gyr$^{-1}]$\\ 
\hline\hline
Miranda & $2.2\times 10^3$ & $9 \times 10^{-4}$ & $7 \times 10^{-5}$ & $2.7\times 10^{-3}$\\
\hline
Ariel & $6.4 \times 10^3$ & $1 \times 10^{-2}$ & $8 \times 10^{-4}$ & $4.6\times 10^{-3}$\\
\hline
Umbriel & $4.5 \times 10^4$ & $7 \times 10^{-3}$ & $6 \times 10^{-4}$ & $4.6 \times 10^{-4}$\\
\hline
Titania & $7.6 \times 10^5$ & $2 \times 10^{-2}$ & $2 \times 10^{-3}$ & $5.5\times 10^{-5}$\\
\hline
Oberon & $5.1 \times 10^6$ & $2 \times 10^{-2}$ & $1 \times 10^{-3}$ & $7.2\times 10^{-6}$\\
\hline
\end{tabular}
\end{center}
\end{table}

Early in this project, we settled on a nominal Uranian tidal $Q_p=40,000$ since it was immediately clear that Miranda's inclination could be generated in the Ariel-Umbriel 5:3 MMR, meaning that the Miranda-Umbriel 3:1 MMR may not have happened. Miranda would likely accumulate excessive inclination if the moons crossed both of these resonances, especially if the Miranda-Umbriel 3:1 MMR was not broken at moderate $i_M$ by a secondary resonance \citep{tw2, mal90}. As \citet{pea88} found that the Miranda-Umbriel 3:1 MMR is crossed if $Q_p < 39,000$, we took $Q_p=40,000$, the smallest tidal $Q$ of Uranus for which that resonance is not crossed, as a good reference value for tidal $Q$ of Uranus.    

Our results completely change the above argument about Uranian $Q_p$ based on resonance crossing. First of all, the calculation in \citet{pea88} which finds $Q_p=39,000$ as critical for Miranda-Umbriel 3:1 MMR crossing was based on a now-outdated mass of Miranda. By the same method we used to generate Fig. \ref{peale}, we calculate that $Q_p \simeq 25,000$ puts this resonance at about $4.5$~Gyr ago. Second, we find that the time the system spends in resonances that enhance the tidal evolution of Umbriel (hundreds of Myr in the Ariel-Umbriel 5:3 MMR and an unknown amount of time in the preceding three-body resonance) further lowers the tidal $Q_p$ for which the system can avoid the Miranda-Umbriel 3:1 MMR. A Uranian tidal lag angle $\delta \approx Q_p^{-1}$ directly proportional to synodic frequency (as opposed to a constant one used here) would also push 3:1 Miranda-Umbriel MMR further to the past relative to Ariel-Umbriel resonances. However, by far the most important change to our understanding of past resonance crossings comes from the shrinking of Miranda's orbit during the Ariel-Umbriel 5:3 MMR. Since this MMR likely lasted hundreds of Myr, Miranda's orbit should have contracted by several percent, with tens of percent not being impossible if the highest dissipation rates were present throughout the duration of the MMR. Therefore, before the Ariel-Umbriel 5:3 MMR, Miranda was likely orbiting much further outside the location of the 3:1 resonance with Umbriel. So Miranda's current near-commensurability with Umbriel does not imply that their resonance would have been crossed in the past during Miranda's outward tidal migration. 

A more straightforward way of estimating Uranus's tidal $Q$ is by connecting the cratering ages of the moons' surfaces to the past dynamical events such as resonance crossings. We already proposed that the young coronae on Miranda, Arden and Inverness, probably formed during the Ariel-Umbriel resonance crossing. The crater-derived age of these features depends on what assumption is made about the size-frequency distribution of cometary impactors \citep{zah03}. Recent results from the New Horizons mission \citep{sin19} clearly indicate that cometary impactors have a shallower size-frequency distribution (i.e. one with fewer small members) than most prior models predicted. Therefore, we will adopt \citet{zah03} ``Case A'' model (which assumes fewer impactors at small sizes) to estimate the age of the Uranian satellite's surfaces. This yields an age of about 1~Gyr (or a little less) for the Arden and Inverness coronae \citep{zah03}.

Given the tidal evolution rates of Ariel and Umbriel as a function of Uranus's tidal $Q$ (Table \ref{tidal}), and the fact that currently we have $5 n_U- 3 n_A = \Delta{n} = 2 \pi (62{\rm d})^{-1}$ (where $n$ is mean motion). For a very recent resonance like Ariel-Umbriel 5:3 MMR involving only 1\% change in Ariel's $a$, a linear estimate of age from current tidal evolution rates is accurate enough to be useful:
\begin{equation}
T {\rm[Gyr]} \approx \Bigl({Q_p \over 40,000}\Bigr) {2 \Delta{n} \over 9 n_A} \Bigl({\dot{a_A} \over a_A} - {\dot{a_U} \over a_U}\Bigr)^{-1} 
\label{reviewer}
\end{equation}
where the $\dot{a}$ are calculated for $Q_p=40,000$ (Table \ref{tidal}), and we assumed $n_U/n_A \approx 3/5$. 
So we can say that the Ariel-Umbriel 5:3 resonance would have happened 2.2~Gyr ago for our nominal $Q_p=40,000$, and the resonance breaking at 1~Gyr ago requires $15,000 < Q_p < 20,000$. Further work on the modeling of heliocentric impacts on Uranian satellites could help improve on this estimate, and better determination of the masses of Ariel and Umbriel should further reduce uncertainty. This value of tidal $Q$ of Uranus is consistent with the Ariel-Umbriel 2:1 MMR having never been crossed, but further work is necessary to establish this. One should keep in mind that the cratering age uncertainties are significant, and also there is no guarantee that the tidal properties of Uranus did not change over time.

\subsection{Tidal Response of the Uranian Moons} \label{subsec:moons}

Having an estimate of the absolute age of the Ariel-Umbriel 5:3 MMR crossing enables us to say more about the tidal properties of the moons. Miranda's current eccentricity of $e_M=0.0013$ \citep{jac14} either requires very fast freezing of Miranda once the Ariel-Umbriel resonance is broken, as Miranda's $Q/k_2=10^3$ required for sufficient tidal heating also gives an eccentricity damping timescale of a couple Myr (Table \ref{tidal}). Miranda would have to transition to a rigid-body-like $Q/k_2=10^5-10^6$ (Table \ref{tidal}; we assume $Q=100$ for the moons in absence of better information) in a few Myr in order to preserve some of the eccentricity from the resonance. Alternatively, Miranda may have obtained eccentricity more recently, through the same three-body resonance as Ariel (Fig. \ref{fish4}); more thorough examination of this resonance is needed, especially with very low pre-resonance eccentricities expected from tidal damping.   

Ariel, or more correctly, secular mode E2, usually has a low eccentricity ($e_2=0.001-0.002$) at the end of the secular resonance with Umbriel shown in Fig. \ref{sec2}. As the mode's current amplitude is $e_2=0.001$, survival of ancient eccentricity would require $Q/k_2 > 10^5$ for Ariel, which would make Ariel exceptionally non-dissipative. Fortunately, the three-body resonance with Umbriel and Titania (Fig. \ref{fish4}), which has an age that is only 30\% of that of the 5:3 MMR with Umbriel, significantly relaxes this constraint. $10^4 < Q/k_2 < 10^5$ for Ariel would result in a tidal damping timescale of a couple hundred Myr, with about one damping timescale since the three-body resonance, so tidal damping could reconcile the simulation in Fig. \ref{fish4} with Ariel's current eccentricity. The same $10^4 < Q/k_2 < 10^5$ for Umbriel would result in $\tau_e > 1$~Gyr, which is in agreement with a substantial fraction of Umbriel's (i.e. mode E3) eccentricity surviving from the MMR (and the subsequent secular resonance) with Ariel. 

In Fig. \ref{oberon8} we show that $Q/k_2=10^4$ for Oberon can reproduce the amplitude of the secular inclination mode I3. Caveats include possible subsequent action of the Ariel-Umbriel secular resonance, and the dependence of spin-orbit resonance breaking on the obliquity of Oberon forced by the resonance, which in turn is a function of the exact masses of the satellites. In any case, just like in cases of Ariel and Umbriel, the rough estimate of Oberon's tidal Love number $k_2$ (with the admittedly naive assumption of $Q=100$) is firmly in between the values expected for ice and rock bodies that are both uniform and rigid, using the approach of \citet{md99}. Therefore, our results suggest that Ariel, Umbriel and Oberon have been fully solid bodies during and after the Ariel-Umbriel 5:3 MMR, or, in absolute terms, for longer than a Gyr. This agrees with the calculation of \citet{pet15} that the 5:3 MMR with Umbriel could not have mobilized ice on Ariel, and that Umbriel and Oberon would have experience even less heating. Miranda, in contrast, must have been at least partially molten during the Ariel-Umbriel 5:3 MMR passage, but is also likely to be fully solid by now. 

Titania offers us fewer direct dynamical constraints, as its tidal dissipation is relatively slow but not necessarily negligible. Unlike the inner three moons which have eccentricities dominated by a single secular mode, Titania and Oberon share secular modes E4 and E5, and any damping of those two modes is likely to be driven primarily by Titania's satellite tides (Table \ref{tidal}). Figure \ref{modes} suggests that the amplitudes of models E4 and E5 are a few times higher at the end of Ariel-Umbriel 5:3 MMR than they are now, potentially indicating some degree of damping. Table \ref{tidal} suggests that the eccentricity damping timescale for Titania can be $\tau_e < 1$~Gyr only for $Q/k_2=10^3$, which is significantly lower than what we estimated for Oberon on the basis of its spin-orbit resonance. This may indicate that the response of Titania to Uranus's tides is in excess of that of a rigid body, and that Titania may have an internal ocean. A future mission to the Uranian system may be the only way to conclusively answer this question. 

\section{Conclusions and Future Work} \label{sec:con}

In this paper we used direct numerical integration to study the past Ariel-Umbriel 5:3 mean-motion resonance, as well as subsequent secular, spin-orbit and three-body resonances. We find that prior analytical work, such as \citet{tw1}, was unable to fully capture the richness of dynamics of Uranian moons, which only becomes evident in a full numerical simulation. The Ariel-Umbriel 5:3 resonance is strongly chaotic and eventually breaks when the eccentricities of the two moons reach $e=0.02-0.03$ range. 

Our most important finding is that Miranda was strongly affected by the Ariel-Umbriel resonance, despite not being a direct participant, due to the strongly coupled secular dynamics of the Uranian moons \citep{las87}. Chaotic secular interactions with Ariel while it is in the MMR lead to excitation of both the eccentricity and inclination of Miranda. The excitation of Miranda's eccentricity is the likely cause of the most recent geological activity on Miranda, including the formation of Arden and Inverness coronae \citep{zah03, bed15}. The excitation of inclination typically leaves Miranda with an orbital tilt of several degrees, naturally explaining its current $4.3^{\circ}$ inclination. 

After the Ariel-Umbriel 5:3 MMR is broken, the system goes through a sequence of secular resonances among the moons' secular modes. These secular resonances transfer eccentricity and inclination from the orbits of Ariel, Titania and Oberon to that of Umbriel. We also find that the expected equilibrium shape of Oberon places the rate of its axial precession very close to a spin-orbit resonance with the secular mode I3, associated with the orbital tilt of Umbriel. This secular spin-orbit resonance is excited during the Ariel-Umbriel 5:3 MMR, and may continue once the MMR is broken. We find that joint action of secular resonances and Oberon's precessional spin-orbit resonance can in principle explain the pattern of orbital inclinations of the four largest moons of Uranus. 

More recently, the system went through a three-body resonance involving Ariel, Umbriel and Titania. We propose that this resonance may have generated the current small eccentricities of Ariel and, possibly, Miranda. We find that there were other three-body resonances in the system's past, which can make it more difficult to reconstruct the system's dynamical history.

During the tidal heating of Miranda in the Ariel-Umbriel 5:3 MMR, Miranda's semimajor axis experiences significant contraction. As a consequence, Miranda and Umbriel are unlikely to have crossed their mutual 3:1 resonance in the past, and the Ariel-Umbriel 5:3 MMR may be the sole source of Miranda's inclination. Based on the published $\simeq 1$~Gyr cratering ages of the young coronae on Miranda, we adopt this age for the Ariel-Umbriel 5:3 MMR crossing, and estimate the tidal $Q$ of Uranus to be about $15,000 < Q < 20,000$ \citep[assuming $k_2=0.1$; ][]{gav77}.  

Based on their current eccentricities and likely past resonant dynamics, we find that Ariel, Umbriel, and Oberon behave as rigid bodies, and are unlikely to have had internal oceans during the Ariel-Umbriel 5:3 MMR crossing. Our results allow, and may even require, Titania to have (or have had) a subsurface ocean. If it exists, this ocean would not be a product of the resonances we studied but would need to have been present beforehand.    

Older coronae on Miranda, as well as non-heavily cratered terrains on Ariel, appear to predate the Ariel-Umbriel 5:3 MMR, indicating a separate, more ancient dynamical event. This is likely to be a secular resonance of some kind acting on Ariel, as mean-motion resonances among Uranian moons do not have the power to cause large-scale melting \citep[cf. ][]{pet15}. Our results suggest that an excitation of Ariel's orbit would likely be passed on to Miranda, removing the need for separate heating events for Miranda. We hope that this paper shows that the dynamics of the Uranian system is far more complex that we previously thought, and that more work is needed to fully understand the orbital and geological history of the moons of Uranus.  

\acknowledgments

M\'C and MEM are supported by NASA Solar System Workings Program award 80NSSC19K0544. We wish to thank Chloe Beddingfield, Richard Cartwright, and Francis Nimmo for helpful discussions. Suggestions by two anonymous reviewers have significantly improved the manuscript.

\bibliography{refs}{}
\bibliographystyle{aasjournal}

\end{document}